\documentclass[aoas,preprint,english]{imsart}

\RequirePackage[OT1]{fontenc}
\RequirePackage{amsthm,amsmath}
\RequirePackage[authoryear]{natbib}
\RequirePackage[colorlinks,citecolor=blue,urlcolor=blue]{hyperref}

\RequirePackage{multirow}
\RequirePackage{graphicx}
\RequirePackage[labelformat=empty]{subfig}
\RequirePackage{mathtools}
\RequirePackage{babel}
\RequirePackage{mathrsfs}
\RequirePackage{dsfont}

\arxiv{aa:aaaa}

\startlocaldefs
\numberwithin{equation}{section}
\theoremstyle{plain}

\endlocaldefs

\begin{document}

\begin{frontmatter}
\title{Logistic Biplots for Ordinal Data with an Application to Job Satisfaction of Doctorate Degree Holders in Spain
}
\runtitle{Biplots for Ordinal Data}

\begin{aug}

\author{\snm{Jos\'e Luis Vicente-Villard\'on}
\ead[label=jlvv]{villardon@usal.es}}
\and
\author{\snm{Julio C\'esar Hern\'andez S\'anchez}\ead[label=jchs]{juliocesar.hernandez.sanchez@ine.es}}

\affiliation{Universidad de Salamanca and Spanish Statistical Office}

\address{Jos\'e Luis Vicente-Villard\'on\\
Universidad of Salamanca\\
Salamanca\\
Spain\\
Julio C\'esar Hern\'andez S\'anchez\\
Spanish Statistical Office\\
Zamora\\
Spain\\
\printead{jlvv}\\
\phantom{E-mail:\ }\printead*{jchs}}

\end{aug}

\begin{abstract}

Biplot Methods allow for the simultaneous representation of individuals and variables of a data matrix. For Binary or Nominal data, Logistic biplots have been recently developed to extend the classical linear representations for continuous data. When data are ordinal, linear, binary or nominal logistic biplots are not adequate and techniques as Categorical Principal Component Analysis (CATPCA) or Item Response Theory (IRT) for ordinal items should be used instead. 

In this paper we extend the Biplot to ordinal data. The resulting method is termed Ordinal Logistic Biplot (OLB). Row scores are computed to have ordinal logistic responses along the dimensions and column parameters produce logistic response surfaces that, projected onto the space spanned by the row scores, define a linear biplot. A proportional odds model is used, obtaining a multidimensional model known as graded response model in the Item Response Theory literature. We study the geometry of such a representation and construct computational algorithms for the estimation of parameters and the calculation of prediction directions. Ordinal Logistic Biplots extend both CATPCA and IRT in the sense that gives a graphical representation for IRT similar to the biplot for CATPCA.

The main theoretical results are applied to the study of job satisfaction of doctorate (PhD) holders in Spain. Holders of doctorate degrees or other research qualifications are crucial to the creation, commercialization and dissemination of knowledge and to innovation. The proposed methods are used to extract useful information from the Spanish data from  the international 'Survey on the careers of doctorate holders (CDH)', jointly carried out Eurostat, the Organisation for Economic Co-operation and Development (OECD) and UNESCO's Institute for Statistics (UIS).
\end{abstract}


\end{frontmatter}

\section{Introduction}

The Biplot method (\cite{Gabriel71}, \cite{Gower-96}) is becoming one of the most popular techniques for analysing multivariate data. Biplot methods are techniques for simultaneous representation of the $I$ rows and $J$ columns of a data matrix ${\bf{X}}$, in reduced dimensions, where rows represent to individuals, objects or samples and columns to variables measured on them. Classical Biplot methods are a graphical representation of a Principal Components Analysis (PCA) or Factor Analysis (FA) that it is used to obtain linear combinations that successively maximize the total variability. From another point of view, Classical Biplots can be obtained from alternated regressions and calibrations  \citep{GabrielZamir1979}. This approach is essentially an alternated least squares algorithm equivalent to an EM-algorithm when data are normal.

For data with distributions from the exponential family, \cite{Gabriel1998}, describes ``bilinear regression'' as a method to estimate biplot parameters, but the procedure have never been implemented and the geometrical properties of the resulting representations have never been studied. \cite{deLeeuw06} proposes Principal Components Analysis for Binary data based on an alternate procedure in which each iteration is performed using iterative majorization and \cite{Lee10} extends the procedure for sparse data matrices, none of those describe the associated biplot. \cite{Vicente-06} propose a biplot based on logistic responses called ``Logistic Biplot'' that is linear, the paper studies the geometry of this kind of biplots and uses a estimation procedure that is slightly different from Gabriel's method. A heuristic version of the procedure for large data matrices in which scores for individuals are calculated with an external procedure as Principal Coordinates Analysis is described in \cite{Demeyetal}.  Method is called ``External Logistic Biplot''. Binary Logistic Biplots have been successfully applied to different data sets, see for example, \cite{GallegoAlvarez2012250}, \cite{Galindo20113} or \cite{Demeyetal}. 
For nominal data,  \cite{NominalBiplot} propose a biplot representation based on convex prediction regions for each category of a nominal variable. EM algorithm is used for the parameter estimation. In section 2, biplots for continuous, binary and, in lesser extent, nominal variables are described.

When data are ordinal, linear, binary or nominal logistic biplots are not adequate and techniques as Categorical Principal Component Analysis (CATPCA) or Item Response Theory (IRT) for ordinal items should be used instead. In Section 3 we extend the Biplot to ordinal data. The resulting method is termed Ordinal Logistic Biplot (OLB). Row scores are computed to have ordinal logistic responses along the dimensions and column parameters produce logistic response surfaces that, projected onto the space spanned by the row scores, define a linear biplot. A proportional odds model is used, obtaining a multidimensional model known as graded response model in the Item Response Theory literature. We study the geometry of such a representation and construct computational algorithms for the estimation of parameters and the calculation of prediction directions. Ordinal Logistic Biplots extend both CATPCA and IRT in the sense that gives a graphical representation for IRT similar to the biplot for CATPCA.
In Section 4 the main results are applied to the study of job satisfaction of doctorate (PhD) holders in Spain. Holders of doctorate degrees or other research qualifications are crucial to the creation, commercialization and dissemination of knowledge and to innovation. The proposed methods are used to extract useful information from the Spanish data from  the international 'Survey on the careers of doctorate holders (CDH)', jointly carried out Eurostat, the Organisation for Economic Co-operation and Development (OECD) and UNESCO's Institute for Statistics (UIS).
Finally, in Section 5, there is a brief discussion of the main results concerning both, statistical and applied.

\section{Logistic Biplot for Continuous, Binary or Nominal Data}

In this section we describe the biplot for continuous, binary and nominal data, being the first two treated in a greater extent because of the closer relation to the proposal of this paper.

\subsection{Continuous Data}
Let ${{\bf{X}}_{I \times J}}$  be a data matrix of continuous measures, and consider the following reduced rank model (S-dimensional)
\begin{equation}
{\bf{X}} = {{\bf{1}}_I}{{\bf{b'}}_0} + {\bf{AB'}} + {\bf{E}}
\label{biplot}
\end{equation}
where ${{\bf{b'}}_0}$ is a vector of constants,  usually the column means(${{\bf{b'}}_0}={\bf{\bar x'}}$), $\bf{A}$ and $\bf{B}$ are matrices of rank $S$ with $I$ and $J$ columns respectively, and $\bf{E}$ is an $I \times J$ matrix of errors. The reduced rank approximation of the centred data matrix (expected values), written as 
\begin{equation}
{\bf{\tilde X}} = E\left[ {{\bf{X}} - {{\bf{1}}_I}{{\bf{b'}}_0}} \right] = {\bf{AB'}}
\label{biplot2}
\end{equation}

or

\begin{equation}
E\left[ {\bf{X}} \right] = {{\bf{1}}_I}{{\bf{b'}}_0} + {\bf{AB'}},
\label{biplot3}
\end{equation}

 is usually obtained from its Singular Value Decomposition (SVD), is closely related to its Principal Components and its called a Biplot \citep{Gabriel71} because it can be used to simultaneously plot the individuals and variables using the rows of ${\bf{A}} = ({{\bf{a}}_1}, \ldots ,{{\bf{a}}_I})'$ and ${\bf{B}} = ({{\bf{b}}_1}, \ldots ,{{\bf{b}}_J})'$ as markers, in such a way that the inner product ${{\bf{a'}}_i}{{\bf{b}}_j}$ approximates the element ${{\tilde x}_{ij}}$ as close as possible.

If we consider the row markers $\bf{A}$ as fixed and the data matrix previously centred, the column markers can be computed by regression trough the origin: 
\begin{equation}
{\bf{B'}} = {({\bf{A'A}})^{ - 1}}{\bf{A'}}({\bf{X}} - {{\bf{1}}_I}{\bf{\bar x'}}).
\label{reg1}
\end{equation}
In the same way, fixing $\bf{B}$, $\bf{A}$ can be obtained as:
\begin{equation}
{\bf{A'}} = {({\bf{B'B}})^{ - 1}}{\bf{B'}}({\bf{X}} - {{\bf{1}}_I}{\bf{\bar x'}})'.
\label{reg2}
\end{equation}
Alternating the steps (\ref{reg1}) and (\ref{reg2}) the product converges to the Singular Value Decomposition (SVD) of the centred data matrix. Pre
Regression step in equation (\ref{reg1}) adjusts a separate linear regression for each column (variable) and interpolation step in equation (\ref{reg2}) interpolates an individual using the column markers as reference. The procedure is in some way a kind of EM-algorithm in which the regression step is the maximization part and the interpolation step is the expectation part. In summary, the expected values on the original data matrix are obtained on the biplot using a simple scalar product, that is, projecting the point ${{\bf{a}}_i}$ onto the direction defined by ${{\bf{b}}_j}$. This is why row markers are usually represented as points and column markers as vectors (also called biplot axis by \cite{Gower-96}). 

\begin{figure}[!htb]
 \centering
 \includegraphics[width=0.7\textwidth]{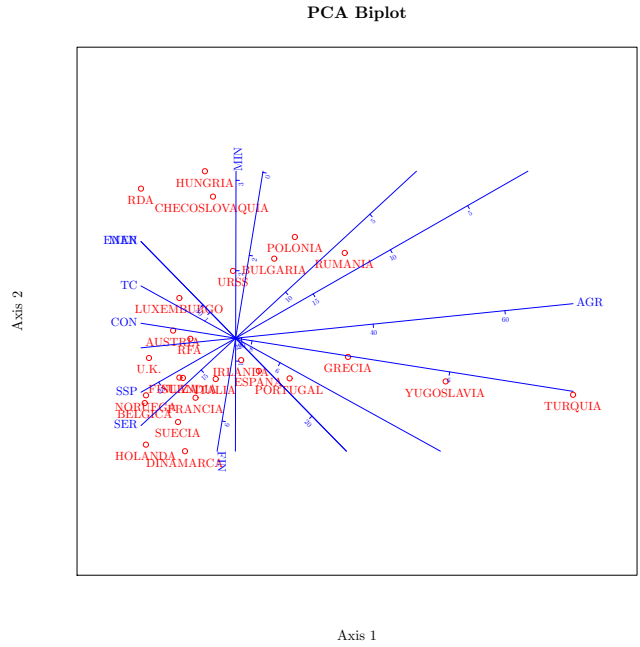}
 \caption{PCA Biplot with scales for the variables.}
 \label{Continuous}     
\end{figure}

The biplot axis can be completed with scales to predict individual values of the data matrix. To find the point on the biplot direction, that predicts a fixed value $\mu $ of the observed variable when an individual point is projected, we look for the point $(x,y)$ lies on the biplot axis, i. e.  that verifies  $$y = {{{b_{j2}}} \over {{b_{j1}}}}x$$
      and      
$$\mu  = {b_{j0}} + {b_{j1}}x + {b_{j2}}y$$
Solving for x and y, we obtain
$$x = (\mu - {b_{j0}}) {{{b_{j1}}} \over {b_{j1}^2 + b_{j2}^2}}$$
and 
$$y = (\mu - {b_{j0}}) {{{b_{j2}}} \over {b_{j1}^2 + b_{j2}^2}}$$
Or
$$(x,y) =  (\mu - {b_{j0}}) {{{{\bf{b}}_j}} \over {{{{\bf{b'}}}_j}{{\bf{b}}_j}}}$$
Therefore, the unit marker for the j-th variable is computed by dividing the coordinates of its corresponding marker by its squared length, several points for specific values of $\mu $ can be labelled to obtain a reference scale. If data are centred, ${b_{j0}}=0 $, labels can be calculated by adding the average to the value of $\mu $ ($\mu  + {{\bar x}_j}$). The resulting representation will be like the one shown in figure \ref{Continuous}

\subsection{Binary Data}
Let ${{\bf{P}}_{I \times J}}$  be a binary data matrix. Let ${\pi_{ij}} = E({x_{ij}})$ the expected probability that the character $j$ be present at individual $i$, and ${p_{ij}}$ the observed probability, either 0 or 1, resulting in a binary data matrix. The S-dimensional logistic biplot  in the $logit $ scale is formulated as:
\begin{equation}
logit({\pi_{ij}}) = \log ({{{\pi_{ij}}} \over {1 - {\pi_{ij}}}}) = {b_{j0}} + \sum\limits_{s = 1}^S {{b_{js}}{a_{is}}}  = {b_{j0}} + {{\bf{a'}}_i}{{\bf{b}}_j},
\label{binlogbip}
\end{equation}
where ${a_{is}}$  and ${b_{js}}$, $(i=1, \dots ,I; j=1, \dots ,J; s=1, ..., S)$, are the model parameters used as row and column markers respectively. The model is a generalized (bi)linear model having the $logit$ as a link function. In terms of probabilities rather than $logits$:
\begin{equation}
{\pi _{ij}} = {{{e^{{b_{j0}} + \sum\nolimits_k {{b_{jk}}{a_{ik}}} }}} \over {1 + {e^{{b_{j0}} + \sum\nolimits_k {{b_{jk}}{a_{ik}}} }}}} = {1 \over {1 + {e^{ - ({b_{j0}} + \sum\nolimits_k {{b_{jk}}{a_{ik}}} )}}}}.
\end{equation}
 In matrix form:  
  \begin{equation}
 logit({\bf{\Pi}}) = {{\bf{1}}_I}{{\bf{b'}}_0} + {\bf{AB'}},
 \label{binlogbip3}
 \end{equation}
where $\bf{\Pi}$ is the matrix of expected probabilities, ${{\bf{1}}_I}$ is a vector of ones and ${{\bf{b}}_0} = ({b_{10}}, \ldots ,{b_{J0}})$  is the vector containing intercepts that have been added because it is not possible to centre the data matrix in the same way as in linear biplots.
 The points predicting different probabilities are on parallel straight lines on the biplot; this means that predictions on the logistic biplot are made in the same way as on the linear biplots, i. e., projecting a row marker ${{\bf{a}}_i} = ({a_{i1}},{a_{i2}})$ onto a column marker ${{\bf{b}}_j} = ({b_{j1}},{b_{j2}})$.  (See \cite{Vicente-06} or \cite{Demeyetal}). The calculations for obtaining the scale markers are simple. To find the marker for a fixed probability $\pi$, we look for the point $(x,y)$ that predicts $\pi$ and is on the ${{\bf{b}}_j}$ direction, i. e., on the line joining the points (0, 0) and $({b_{j1}},{b_{j2}})$, that is
 $$y = {{{b_{j2}}} \over {{b_{j1}}}}x$$
 and $$logit(\pi ) = {b_{j0}} + {b_{j1}}x + {b_{j2}}y$$
 
 We obtain
 $$x = {{(logit(\pi ) - {b_{j0}}){b_{j1}}} \over {b_{j1}^2 + b_{j2}^2}}$$
 and
 $$y = {{(logit(\pi ) - {b_{j0}}){b_{j2}}} \over {b_{j1}^2 + b_{j2}^2}}$$
 several points for specific values of $\pi $ can be labelled to obtain a reference scale. From a practical point of view the most interesting value is  $\pi  = 0.5$ because the line passing trough that point and perpendicular to the direction divides the representation into two regions, one that predicts \emph{presence} and other \emph{absence}. Plotting that point and an arrow pointing to the direction of increasing probabilities should be enough for most practical applications.

The model in (\ref{binlogbip}) is also a latent trait or item response theory model, in that ordination axes are considered as latent variables that explain the association between the observed variables. In this framework we suppose that individuals respond independently to variables, and that the variables are independent for given values of the latent traits.
With these assumptions the likelihood function is:
 \begin{equation}
 {\rm{Prob}}(\left. {{x_{ij}}} \right|({{\bf{b}}_0},{\bf{A}},{\bf{B}})) = \prod\limits_{i = 1}^I {\prod\limits_{j = 1}^J {\pi _{ij}^{{x_{ij}}}{{(1 - {\pi _{ij}})}^{1 - {x_{ij}}}}} }.
 \label{binlikelyhood}
 \end{equation}
Taking the logarithm of the likelihood function yields:
 \begin{equation}
 L = {\rm{log}}\;{\rm{Prob}}(\left. {{x_{ij}}} \right|({{\bf{b}}_0},{\bf{A}},{\bf{B}})) = \sum\limits_{i = 1}^I {\sum\limits_{j = 1}^J {\left[ {{x_{ij}}\log ({\pi _{ij}}) + (1 - {x_{ij}})\log (1 - {\pi _{ij}})} \right]} }.
  \label{binloglikelyhood}
  \end{equation}
  
For $\bf{A}$ fixed, (\ref{binloglikelyhood}) can be separated into $J$ parts, one for each variable:
   \begin{equation}
   L = \sum\limits_{J = 1}^J {{L_j}}  = \sum\limits_{J = 1}^J {\left( {\sum\limits_{i = 1}^I {\left[ {{x_{ij}}log({\pi _{ij}}) + (1 - {x_{ij}})log(1 - {\pi _{ij}})} \right]} } \right)}.
  \end{equation}
Maximizing each ${{L_j}}$ is equivalent to performing a standard logistic regression using the $j$-$th$ column of $\bf{X}$ as a response and the columns of $\bf{A}$ as regressors. 
In the same way the probability function can be separated into several parts, one for each row of the data matrix, $L = \sum\nolimits_{i = 1}^I {{L_i}} $. The details of the procedure  to calculate the row or individual scores can be found in \cite{Vicente-06}.
 
Binary logistic biplots can be calculated using the package MULTBIPLOT \cite{MULTBIPLOT}. A typical representation

\subsection{Nominal Data}

Let ${{\bf{X}}_{I \times J}}$ be a data matrix containing the values of $J$ nominal variables, each with $K_j$ $(j=1, \ldots, J)$ categories, for $I$ individuals, and let ${{\bf{P}}_{I \times L}}$  be the corresponding indicator matrix with  $L = \sum\nolimits_j {{K_j}} $ columns.  The last (or the first) category of each variable will be used as a baseline. Let ${\pi _{ij(k)}}$ denote the expected probability that the category $k$ of variable $j$ be present at individual $i$. 
A multinomial logistic latent trait model with $S$ latent traits, states that the probabilities are obtained as:
\begin{equation}
{\pi _{ij(k)}} = {{{e^{{b_{j(k)0}} + \sum\limits_{s = 1}^S {{b_{j(k)s}}{a_{is}}} }}} \over {\sum\limits_{l = 1}^{{K_j}} {{e^{{b_{j(l)0}} + \sum\limits_{s = 1}^S {{b_{j(l)s}}{a_{is}}} }}} }}, (k = 1, \ldots ,{K_j}).
  \label{NominalProb}
\end{equation}
Using the last category as a baseline in order to make the model identifiable, the parameter for that category are restricted to be 0, i.e., ${b_{j({K_j})0}} = {b_{j({K_j})s}} = 0$, $(j = 1, \ldots ,J;\quad s = 1, \ldots ,S)$.The model can be rewritten as: 
\begin{equation}
{\pi _{ij(k)}} = {{{e^{{b_{j(k)0}} + \sum\limits_{s = 1}^S {{b_{j(k)s}}{a_{is}}} }}} \over {1 + \sum\limits_{l = 1}^{{K_j} - 1} {{e^{{b_{j(l)0}} + \sum\limits_{s = 1}^S {{b_{j(l)s}}{a_{is}}} }}} }}, (k = 1, \ldots ,{K_j} - 1).
  \label{NominalProb2}
\end{equation}
With this restriction we assume that the log-odds of each response (relative to the last category) follows a linear model: 
$$\log \left( {{{{\pi _{ij(k)}}} \over {{\pi _{ij({K_j})}}}}} \right) = {b_{j(k)0}} + \sum\limits_{s = 1}^S {{b_{j(k)s}}{a_{is}}}  = {b_{j(k)0}} + {{{\bf{a'}}}_i}{{\bf{b}}_{j(k)}},$$
where ${{a_{is}}}$  and ${b_{j(k)s}}\quad (i = 1, \ldots ,I;\quad j = 1, \ldots ,J;\quad k = 1, \ldots ,{K_j} - 1;\quad s = 1, \ldots ,S)$ are the model parameters. In matrix form:
\begin{equation}
{\bf{O}} = {{\bf{1}}_I}{{{\bf{b'}}}_0} + {\bf{AB'}},
  \label{NominalOdds}
\end{equation}
where ${{\bf{O}}_{I \times (L - J)}}$ is the matrix containing the expected log-odds, defines a biplot for the odds. Although the biplot for the odds may be useful, it would be more interpretable in terms of predicted probabilities and categories. This Biplot will be called ``Nominal Logistic Biplot'', and it is related to the latent nominal models in the same way as classical linear biplots are related to Factor or Principal Components Analysis or Binary Logistic Biplots are related to the Item Reponse Theory or Latent Trait Analysis for Binary data.

The points predicting different probabilities are no longer on parallel straight lines; this means that predictions on the logistic biplot are not made in the same way as in the linear biplots, the response surfaces define now prediction regions for each category as shown in \cite{NominalBiplot}. The Nominal Logistic Biplot is described here in lees detail because its geometry is lees related to our proposal than Linear or Binary Logistic Biplots.

\section{Logistic Biplot for Ordinal Data}
\subsection[GNLB]{Formulation and Geometry}
Let ${{\bf{X}}_{I \times J}}$ be a data matrix containing the measures of $I$ individuals on $J$ ordinal variables with $K_j , (j=1,\dots,J)$ ordered categories each, and let  ${{\bf{P}}_{I \times L}}$ the indicator matrix with $L=\sum_j(K_j)$ columns. The indicator $I \times K_j$ matrix for each categorical variable ${{\bf{P}}_j}$ contains binary indicators for each category and ${\bf{P}} = \left( {{{\bf{P}}_1}, \ldots ,{{\bf{P}}_J}} \right)$.  Each row of $\bf{P}_j$ sums 1 and each row of $\bf{P}$ sums $J$.  Then $\bf{P}$ is the matrix of observed probabilities for each category of each variable.

\begin{figure}[!htb]
 \centering
 \includegraphics[width=0.7\textwidth]{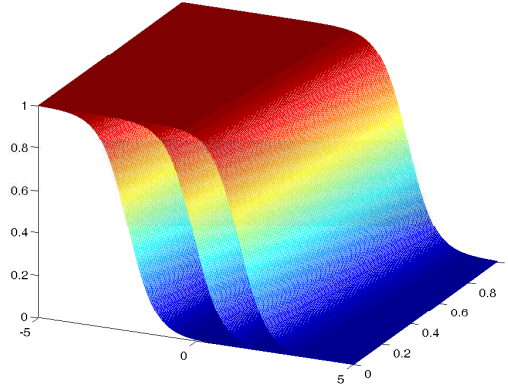}
 \caption{Cumulative response curves for a two-dimensional latent trait and a variable with four categories.}
 \label{cumulative}     
\end{figure}

Let $\pi _{ij(k)}^{*} = P(x_{ij} \leq k)$ be the (expected) cumulative probability that individual $i$ has a value lower than $k$ on the $j-th$ ordinal variable, and let $\pi _{ij(k)} = P(x_{ij} = k)$ the (expected) probability that individual $i$ takes the $k-th$ value on the $j-th$ ordinal variable. Then $\pi _{ij(K_{j})}^{*} = P(x_{ij} = K_{j}) =1 $ and $\pi _{ij(k)} = \pi _{ij(k)}^{*} - \pi _{ij(k-1)}^{*}$ (with $\pi _{ij(0)}^{*}=0$). A multidimensional (S-dimensional) logistic latent trait model for the cumulative probabilities can be written for ($1 \le k \le K_{j}-1$) as
\begin{equation}
\pi _{ij(k)}^* = {1 \over {1 + {e^{ - \left( {{d_{jk}} + \sum\nolimits_{s = 1}^S {{a_{is}}{b_{js}}} } \right)}}}} = {1 \over {1 + {e^{ - ({d_{jk}} + {{{\bf{a'}}}_i}{{\bf{b}}_j})}}}}
 \label{cumulativeprob}
\end{equation}
where ${{\bf{a}}_i} = ({a_{i1}}, \ldots ,{a_{iS}})'$ is the vector of latent trait scores for the $i-th$ individual and ${d_{jk}}$ and ${{\bf{b}}_j} = ({b_{j1}}, \ldots ,{b_{jS}})'$ the parameters for each item or variable. Observe that we have defined a set of binary logistic models, one for each category, where there is a different intercept for each but a common set of slopes for all. In the context of Item Response Theory, this is known as the \emph{Graded Response Model} or \emph{Samejima's Model} \cite{Samejima-69}. The main difference with IRT models is that we don't have the restriction that the probability of obtaining a higher category must increase along the dimensions. Our variables are not necessarily items of a test, but the models are formally the same for both cases. For the unidimensional case that corresponds to a model with an unique discrimination ${b_{j}}$ for all categories and different threshold, boundaries, difficulties or location parameters ${d_{j(k)}}$. The unidimensional cumulative model is shown in Figure \ref{cumulative}.
The ${\bf{a}}_i$ scores can be represented in a scatter diagram and used to establish similarities and differences among individuals or searching for clusters with homogeneous characteristics, i. e., the representation is like the one obtained from any Multidimensional Scaling Method. In the following we will see that the ${{\bf{b}}_j}$ parameters can also be represented on the graph as directions on the scores space that best predict probabilities and are used to help in searching for the variables or items responsible for the differences among individuals. 

In logit scale, the model is
\begin{equation}
logit(\pi_{ij(k)}^{*})=d_{j(k)} + \sum_{s=1}^{S} a_{is} b_{js}=d_{j(k)} + {\bf{a'}}_i{\bf{b}}_j \medspace, \medspace k=1,\dots,K_j-1
\label{clmodel}
\end{equation}
That defines a Binary Logistic Biplot for the cumulative categories.

In matrix form:
\begin{equation}
logit({\Pi ^*}) = {\bf{1d'}} + {\bf{AB'}}
  \label{OrdinalOdds}
\end{equation}
where ${\Pi ^*} = (\Pi _1^*, \ldots ,\Pi _J^*)$ is the $ I \times (L-J)$ matrix of expected cumulative probabilities, ${{\bf{1}}_{I}}$ is a vector of ones and ${\bf{d}} = ({{{\bf{d'}}}_1}, \ldots ,{{{\bf{d'}}}_J})$, with  ${{{\bf{d'}}}_j} = ({d_{j(1)}}, \ldots ,{d_{j({K_j} - 1)}})$,  is the vector containing thresholds, ${\bf{A}} = ({{{\bf{a'}}}_1}, \ldots ,{{{\bf{a'}}}_I})'$ with ${{{\bf{a'}}}_i} = ({a_{i1}}, \ldots ,{a_{iS}})$ is the $ I \times S$ matrix containing the individual scores matrix and ${\bf{B}} = ({{{\bf{B'}}}_1}, \ldots ,{{{\bf{B'}}}_J})'$ with ${{\bf{B}}_j} = {{\bf{1}}_{{K_j} - 1}} \otimes {{{\bf{b'}}}_j}$ and ${{{\bf{b'}}}_j} = ({b_{j1}}, \ldots ,{b_{jS}})$, is the $ (L-J) \times S$ matrix containing the slopes for all the variables. This expression defines a biplot for the odds that will be called ``Ordinal Logistic Biplot''. Each equation of the cumulative biplot shares the geometry described for the binary case \citep{Vicente-06}, moreover, all  curves share the same direction when projected on the biplot. 
The set of parameters $\{d_{jk}\}$ provide a different threshold for each cumulative category, the second part of   (\ref{clmodel}) does not depend on the particular category, meaning that all the $K_j-1$ curves share the same slopes.
In the following paragraphs we will obtain the geometry for the general case an algorithm to perform the calculations.

The expected probability of individual $i$ responding in category $k$ to item $j$, with $(k=1,\dots,K_j)$, that we denote by $\pi_{ij(k)} = P(x_{ij} = k)$
must be obtained by subtracting cumulative probabilities:
$$\pi _{ij(k)} = \pi _{ij(k)}^{*} - \pi _{ij(k-1)}^{*}$$
then using the equations in (\ref{cumulativeprob}):
\begin{equation}
\begin{split}
\pi_{ij(1)}&=P(x_{ij}=1)=\frac{1}{1+e^{-(d_{j1}+{\bf{a'}}_i{\bf{b}}_j)}}\\
\pi_{ij(k)}&=P(x_{ij}=k)=P(x_{ij}\leq k)- P(x_{ij}\leq (k-1))\\
&=\frac{1}{1+e^{-(d_{jk}+{\bf{a'}}_i{\bf{b}}_j)}}-\frac{1}{1+e^{-(d_{j(k-1)}+{\bf{a'}}_i{\bf{b}}_j)}}\\
&=\frac{e^{-({\bf{a'}}_i{\bf{b}}_j)}(e^{-d_{j(k-1)}}-e^{-d_{jk}})}{(1+e^{-(d_{jk}+{\bf{a'}}_i{\bf{b}}_j)})(1+e^{-(d_{j(k-1)}+{\bf{a'}}_i{\bf{b}}_j)})} \thickspace , \thickspace 1<k<K_j\\
\pi_{ij(K_j)}&=P(x_{ij}=K_j)=1-\frac{1}{1+e^{-(d_{j(K_j-1)}+{\bf{a'}}_i{\bf{b}}_j)}}
\end{split}
\label{pisequations}
\end{equation}

\begin{figure}[!htb]
   \centering
   \subfloat[]{\label{fig:b} \includegraphics[width=0.50\textwidth]{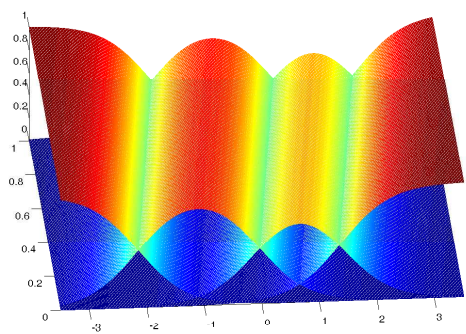}}\\[1pt]
   \caption{Response curves for an ordinal variable with four ordered categories. }
   \label{sigmoides}     
\end{figure}
\begin{figure}[!htb]
 \centering
 \includegraphics[width=0.4\textwidth]{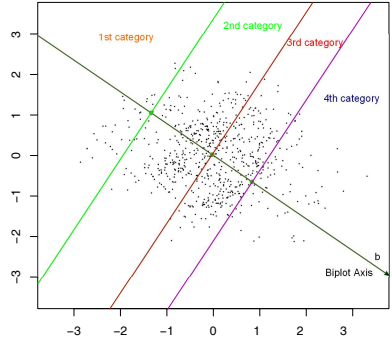}
 \caption{Prediction regions determined by three parallel straight lines for an ordinal variable with four categories.}
 \label{regnivelsat2}     
\end{figure}

If the row scores where known, obtaining the parameters of the model in \ref{pisequations} is equivalent to fitting a proportional odds model using each item as a response and the row scores as regressors. The response surfaces for such a model are shown in figure \ref{sigmoides}. 
Although the response surfaces are no longer sigmoidal, the level curves are still straight lines, so the set of points on the representation (generated by the columns of $ {\bf{A}} $) predicting a particular value for the probability of a category
lie on a straight line, and different probabilities for all the categories of a particular variable or item lie on parallel straight lines. A perpendicular to all those lines  can be used as "biplot axis" as in \cite{Gower-96} and is the direction that better predicts the probabilities of all the categories in the sense that, projecting any individual point onto that direction, we should obtain an optimal prediction of the category probabilities. As all the categories share the same biplot direction, it would be very difficult to place a different graded scale for each and we will represent just the line segments in which the probability of a category is higher then the probability of the rest. That will result, except for some pathological cases, in as many segments as categories ($K_j$), separated by $K_j -1 $ points in which the probabilities of two (contiguous) categories are equal. See figure \ref{regnivelsat2} in which we show the parallel lines representing the points that predict equal probabilities for two contiguous categories and a line, perpendicular to all, that is the "Biplot axis". The three parallel lines divide the space spanned by the columns of ${\bf{A}}$ into four regions, each predicting a particular category of the variable. For a biplot representation we don't need the whole set of lines but just the "axis" and the points on it, intersecting the boundaries of the prediction regions. 

\subsection{Obtaining the biplot representation}
So, if we call $ (x,y) $ one of those intersection points, it must be on the biplot direction, that is, 
\begin{equation}
y = {{{b_{j2}}} \over {{b_{j1}}}}x
\label{slope}
\end{equation}

and the probability of two, possibly contiguous, categories (for example $l$ and $m$) at this point, must be equal, 

\begin{equation}
{\pi _{j(l)}} = {\pi _{j(m)}}\quad (\pi _{j(l)}^* - \pi _{j(l - 1)}^* = \pi _{j(m)}^* - \pi _{j(m - 1)}^*).
\label{equalprob}
\end{equation}
We have omitted index $i$ because probabilities are for a general point and not for a particular individual. Using the condition in \ref{slope} we can rewrite the cumulative probabilities (or its $logit$) as
\begin{equation}
logit(\pi _{j(k)}^*) = {d_{j(k)}} + x{b_{j1}} + y{b_{j2}} = {d_{j(k)}} + z
\label{newlogit}
\end{equation}
with
\begin{equation}
z = x\left( {{{b_{j1}^2 + b_{j2}^2} \over {{b_{j}}}}} \right)
\label{zeta}
\end{equation}
Changing the values of $z$ we can obtain the probabilities of each category along the biplot axis. So, finding the point  $ (x,y) $ is equivalent to find the values of $z$ in which \ref{equalprob} holds. From those values the original point is obtained solving for $x$ in \ref{zeta} and then calculating $y$ from \ref{slope}.

There are some pathological cases in which the probability of one or several categories are never higher than the probability of the rest, in such cases we say that the category is ``hidden'' or ``never predicted'' and the number of separating points will be lower than $K_j -1 $.  Those pathological cases have to be taken into account when calculating the intersection points.

\begin{figure}[!htb]
   \centering
   \subfloat[(a)]{\label{fig:a} \includegraphics[width=0.50\textwidth]{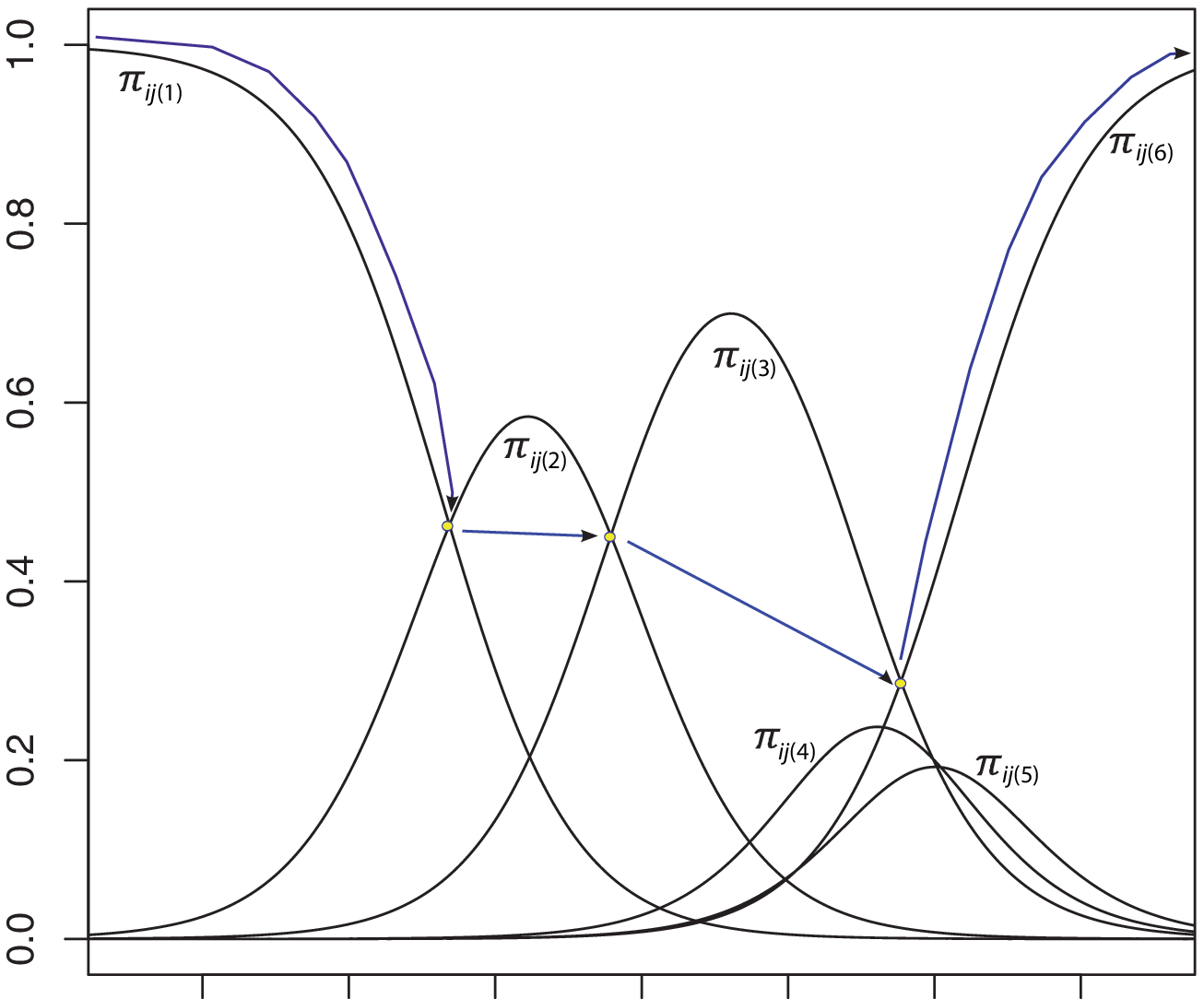}}
   \subfloat[(b)]{\label{fig:b} \includegraphics[width=0.41\textwidth]{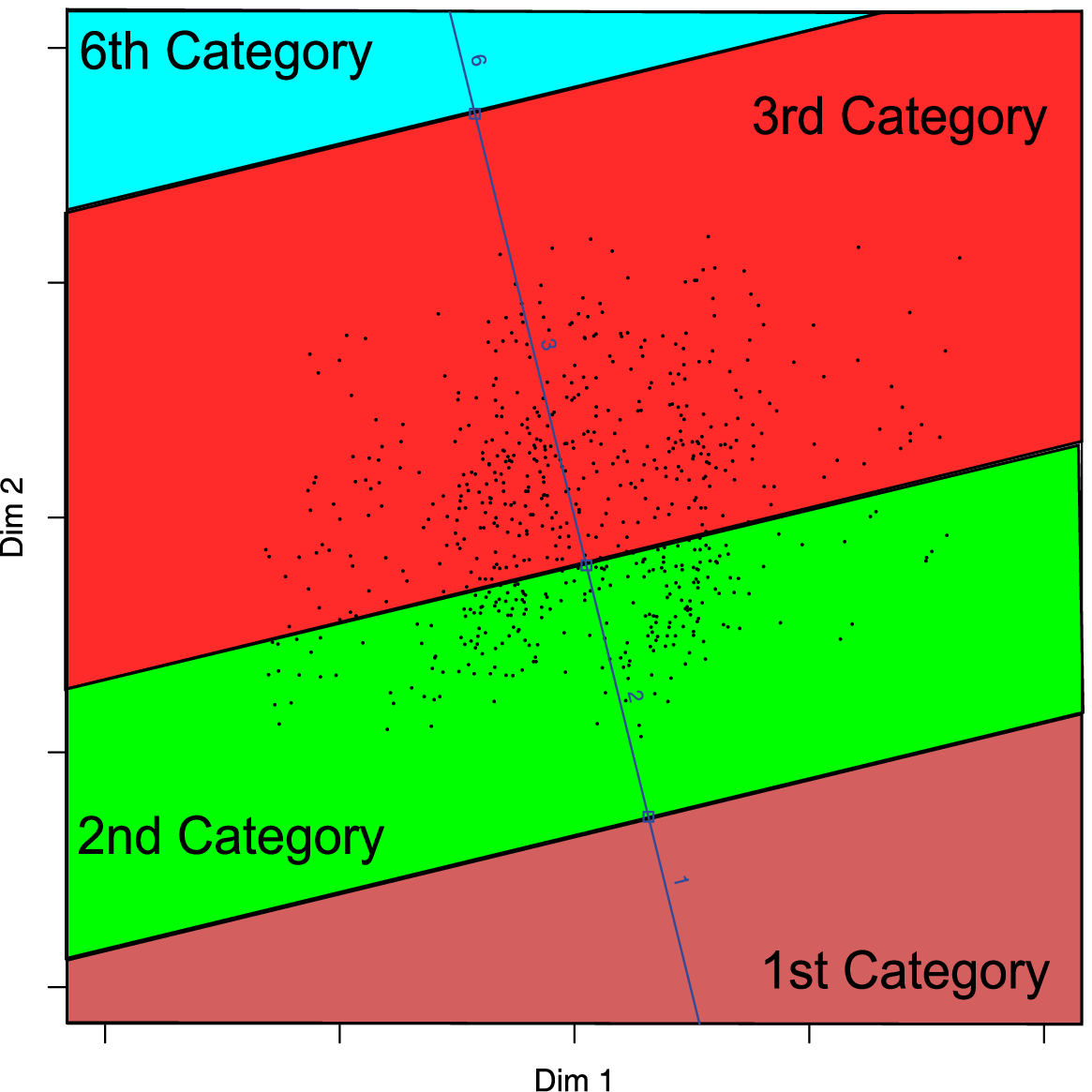}}\\[4pt]
   \caption{Probability curves for a variable with 6 categories in which two (4 and 5) are hidden or never predicted. (a) Projection of the response curves onto a plane prependicular to the biplot axis. (b) Final representation without the hidden categories. }
   \label{hidecat}     
\end{figure}

The existence of abnormal cases means that, not just contiguous, but any pair of categories may have to be compared. Then, many comparisons are possible because the equations are different for each case
\begin{enumerate}
\item $1$-$2$
\item $1$-$l(l<K_j)$
\item $1$-$K_j$
\item $l$-$K_j (l>1)$
\item $l$-$(l+1)$ with $l>1$
\item $l$-$j$ with $j>(l+1), l>1$
\item $(K_j-1)$-$K_j$
\end{enumerate}

For example, in case (3) $1$-$K_j$, is simple to deduce that 
$$z = {{ - ({d_{j({K_j} - 1)}} + {d_{j(1)}})} \over 2}.$$

Cases (1), (3), (5) and (7) are simple. In the other 3 combinations we have to solve a quadratic equation to obtain the intersection points. 
For example, in case (2), the first with the $l-th$ categories, we have to solve  $\pi_{j(1)}=\pi_{j(l)}$, that is
$${1 \over {1 + {e^{ - ({d_{j(1)}} + z)}}}} = {{{e^{ - z}}({e^{ - {d_{j(l - 1)}}}} - {e^{ - {d_{j(l)}}}})} \over {(1 + {e^{ - ({d_{j(l)}} + z)}})(1 + {e^{ - ({d_{j(l - 1)}} + z)}})}}$$
Calling $$w = {e^{ - z}}$$ we have to solve the quadratic equation $$\alpha {w^2} - \beta w - 1 = 0$$ with $\alpha  = ({e^{ - ({d_{j(1)}} + {d_{j(l - 1)}})}} - {e^{ - ({d_{j(1)}} + {d_{j(l)}})}} - {e^{ - ({d_{j(l - 1)}} + {d_{j(l)}})}})$ and $\beta  = 2{e^{ - {d_i}}}$.

If the roots of the equation are both negative, curves don't intersect. If it has a positive root, we can calculate the intersection points solving for $w$ and then reversing the transformations to obtain $(x, y)$.
In a similar way we can calculate the intersection points for cases (4) $i$-$K_j (i>1)$ and (6) $i$-$j$ with $j>(i+1)$.

A procedure to calculate the representation of an ordinal variable on the biplot would be as follows:

\begin{enumerate}
\item Calculate the biplot axis with equation $y = {{{b_{j2}}} \over {{b_{j1}}}}x$.
\item Calculate the intersection points $z$ and then $(x,y)$ of the biplot axis with the parallel lines used as boundaries of the prediction regions for each pair of categories, in the following order: 
\begin{equation*}
\begin{split}
\pi_{j(1)}&=\pi_{j(2)}\\
\pi_{j(l-1)}&=\pi_{j(l)} \thickspace , 1<l<(K_j-1)\\
\pi_{j(K_j-1)}&=\pi_{j(K_j)}
\end{split}
\label{algorithmP}
\end{equation*}

\item If the values of $z$ are ordered, there are not hidden categories and the calculations are finished.

\item If the values of $z$ are not ordered we can do the following:
\begin{enumerate}
\item Calculate the $z$ values for all the pairs of curves, and the probabilities for the two categories involved. 

\item Compare each category with the following, the next to represent is the one with the highest probability at the intersection.

\item If the next category to represent is $K_j$ the process is finished. It not go back to the previous step, starting with the new category.
\end{enumerate}
\end{enumerate}
A simpler algorithm based on a numeric procedure could also be developed to avoid the explicit solution of the equations.

\begin{enumerate}
\item Calculate the predicted category for a set of values for $z$. For example a sequence from -6 to 6 with steps of 0.001. (The precision of the procedure can be changed with the step)
\item Search for the $z$ values in which the prediction changes from one category to another. 
\item Calculate the mean of the two $z$ values obtained in the previous step and then the $(x,y)$ values. Those are the points we are searching for.
\end{enumerate}
Hidden categories are the ones with zero frequencies in the predictions obtained by the algorithm. 

\subsection[GNLB]{Paramater estimation}
The alternated algorithm described in \cite{Vicente-06}, can be easily extended replacing binary logistic regressions by ordinal logistic regressions. The problem with this approach is that the parameters for the individuals can not be estimated when the individual has 0 or 1 in all the variables for the binary case, or all the responses are at the baseline category for the ordinal case. In this paper we use a procedure that is similar to the alternated regressions method, except that the interpolation step is ``changed'' by \emph{a posteriori} expected values. The estimation procedure is an EM-algorithm that uses the Gauss-Hermite quadrature to approximate the integrals, considering the individual scores as missing data.   More details of similar procedures can be found in \cite{Bock81} or \cite{Chalmers2012}.

The likelihood function is:
$$M\left( {\left. {\bf{P}} \right|{\bf{d}},{\bf{A}},{\bf{B}}} \right) = \prod\limits_{i = 1}^I {\prod\limits_{j = 1}^J {\prod\limits_{k = 1}^{{K_j}} {\pi _{ij(k)}^{{p_{ij(k)}}}} } } ,$$
where ${{p_{ij(k)}}}=1$  if individual $i$ chooses category $k$ of item $j$ and ${{p_{ij(k)}}}=0$ otherwise. 
The log-likelihood is:

\begin{equation}
L\left( {\left. {\bf{P}} \right|{{\bf{d}}},{\bf{A}},{\bf{B}}} \right) = \sum\limits_{i = 1}^I {\sum\limits_{j = 1}^J {\sum\limits_{k = 1}^{{K_j}} {{p_{ij(k)}}} } \log \left( {{\pi _{ij(k)}}} \right)}.
\label{LogLikNom}
\end{equation}

If the parameters $\bf{A}$ for  individuals where known, the log-likelihood could be separated into $J$ parts, one for each variable:

\begin{equation}
L\left( {\left. {\bf{P}} \right|{{\bf{d}}},{\bf{B}}} \right) = \sum\limits_{j = 1}^J {{L_j}(\left. {\bf{P}} \right|{{\bf{d}}_{j}},{{\bf{b}}_j})}  = \sum\limits_{j = 1}^J {\left( {\sum\limits_{i = 1}^I {\sum\limits_{k = 1}^{{K_j}} {{p_{ij(k)}}} } \log \left( {{\pi _{ij(k)}}} \right)} \right)},
\label{LogLikNom2}
\end{equation}
where ${{{\bf{d}}_{j}}}$ and ${{{\bf{b}}_j}}$ are the submatrices of parameters for the $j$th variable. Maximizing the log-likelihood is equivalent to maximizing each part, i.e., obtaining the parameters for each variable separately. Maximizing each $L_j$ is equivalent to performing an ordinal logistic regression using the $j$th column of $\bf{X}$ as response and the columns of $\bf{A}$ as predictors. We do not describe logistic regression here because it is as a very well known procedure. It is also well-known that when the individuals for different categories are separated (or quasi-separated) on the space spanned by the explanatory variables, the maximum likelihood estimators don't exist (or are unstable). Because we are seen the biplot as a procedure to classify the set of individuals and searching for the variables responsible for it, accounting for as much of the information as possible, it is probable that, for some variables, the individuals are separated and then the procedure does not work just because the solution is good.   The problem of the existence of the estimators in logistic regression can be seen in \cite{Albert1984}, a solution for the binary case, based on the Firth's method \citep{Firth1993} is proposed by \cite{Heinze2002}. All the procedures were initially developed to remove the bias but work well to avoid the problem of separation. Here we have chosen a simpler solution based on ridge estimators for logistic regression \citep{Cessie1992}.  

Rather than maximizing ${{L_j}(\left. {\bf{P}} \right|{{\bf{d}}_{j}},{{\bf{b}}_j})}$ we maximize:
\begin{equation}
{{L_j}(\left. {\bf{P}} \right|{{\bf{d}}_{j}},{{\bf{b}}_j})} - \lambda \left( {\left\| {{{\bf{d}}_{j}}} \right\| + \left\| {{{\bf{b}}_j}} \right\|} \right).
\label{Penalized}
\end{equation}

We don't describe here the procedure in great detail because that is also a standard procedure. Changing the values of $\lambda$ we obtain slightly different solutions not affected by the separation problem.

In the same way, if  parameters for  variables were known, the log-likelihood could be separated into $I$ parts, one for each individual:
$$L\left( {\left. {\bf{P}} \right|{\bf{A}}} \right) = \sum\limits_{i = 1}^I {{L_i}(\left. {\bf{P}} \right|{{\bf{a}}_i})}  = \sum\limits_{i = 1}^I {\left( {\sum\limits_{j = 1}^J {\sum\limits_{k = 1}^{{K_j}} {{p_{ij(k)}}\log } } \left( {{\pi _{ij(k)}}} \right)} \right)}. $$
To maximize each part we could use Newton-Raphson with  a penalization as before. Rather than that we will use expected a posteriori estimators for the individual markers. 
For each individual (or response pattern) ${{\bf{p}}_i}$, the likelihood is:
$${M}\left( {\left. {\bf{P}}={\bf{p}}_{i} \right|{{\bf{d}}},{\bf{A}}={{\bf{a}}_i},{\bf{B}}} \right) = \prod\limits_{j = 1}^J {\prod\limits_{k = 1}^{{K_j}} {\pi _{ij(k)}^{{p_{ij(k)}}}} }. $$
Assuming a distributional form $g({\bf{a}})$ (multivariate normal, for example) the marginal distribution becomes:
$${P}( {\bf{P}}={\bf{p}}_i | {\bf{d}},{\bf{B}} ) = \int {{M}\left( {\left. {\bf{P}}={{{\bf{p}}_i}} \right|{{\bf{d}}},{\bf{A}}={\bf{a}},{\bf{B}}} \right)} g({\bf{a}})d{\bf{a}},$$
and the observed likelihood:
$$M({\bf{P}} | {\bf{d}},{\bf{B}} ) = \prod\limits_{i = 1}^I \left[ \int {M}({\bf{P}}={\bf{p}}_{i} |{\bf{d}},{\bf{A}}={\bf{a}},{\bf{B}} ) g({\bf{a}})d{\bf{a}} \right]. $$
We approximate the integral by $S$-dimensional Gauss-Hermite quadrature:
\begin{equation}
{{\tilde P}({\bf{P}}={\bf{p}}_i | {\bf{d}},{\bf{B}})} = \sum\limits_{qS = 1}^Q { \ldots \sum\limits_{q1 = 1}^Q {{M}\left( {\left. {\bf{P}}={\bf{p}}_{i} \right|{{\bf{d}}},{\bf{Y}}=y,{\bf{B}}} \right)g({y_{q1}})}  \ldots g({y_{qS}})}.
\label{aproxPL} 
\end{equation}
The multivariate $S$-dimensional quadrature, \textbf{Y}, has been obtained as the product of $S$ unidimensional quadratures $({y_1}, \ldots ,{y_Q})$ with $Q$ nodes each, $\{g(y_q):q=1,\cdots,Q\}$ are associated weights in the quadrature and $y_{q1 \cdots qS}\overset{not}{=}y$ represents each $S$-dimensional quadrature points . Then the marginal expected a posteriori score for each individual can be approximated by:
\begin{equation}
E({\bf{a}}|{\bf{p}}_{i}) \cong {{\sum\limits_{qS = 1}^Q  \ldots \sum\limits_{q1 = 1}^Q {y_{q1 \cdots qS}} P({\bf{P}} = {\bf{p}}_{i} |{\bf{d}},{\bf{Y}}=y_{q1 \cdots qS},{\bf{B}})g({y_{q1}}) \ldots g({y_{qS}})} \over { \tilde{P}({\bf{P}} = {\bf{p}}_{i}| {\bf{d}},{\bf{B}})}},
\label{exapos}
\end{equation}
being $y_{q1 \cdots qS}$ the $S$-dimensional points of the multivariate quadrature, as it was denoted before, and ${{\tilde P}({\bf{P}}={\bf{p}}_i | {\bf{d}},{\bf{B}})}$ given by (\ref{aproxPL}).

The ability for individual $i$ has $S$ components (as much as dimensions of spanned space, i.e. (${\bf{a}}_i=(a_{i1},\cdots,a_{iS})$), and each one $\{a_{is}, s=1,\cdots,S\}$ will be approximated by the expression (\ref{exapos}), which depends on each $S$-dimensional coordinate of $y_{q1 \cdots qS}$.

\section{An empirical study}

In 2008, a set of 26 countries worldwide, among which was Spain, setting 2006 as reference year and following
the guidelines set by the Organization for Cooperation and Development(OECD), the Department of
statistics of UNESCO and Eurostat (Statistical Office of the European Union), began to do surveys to people that had obtained a PhD degree, and that therefore are doctorates, with the objective
of having a clearer information about their characteristics. Most of the pioneer countries belonged 
the European Union,  although members of the OECD, as USA or Australia also participated. In the
Spanish case, it was the National Institute of Statistics (INE) which focused all efforts to carry
out this new  operation with the objective that the availability of information in this field had
continuity in time. Thus, the so-called ``Survey on Human Resources in Science and
Technology'' was established as part of the general plan of science and
technology statistics carried out by the European Union Statistics Office (Eurostat). This need for information is evident in the European Regulation 753/2004 on Science and Technology, which specifies the production of statistics on human resources in science and technology.
 
Surveys on doctorates (CDH: Careers of doctorate holders) try to measure specific demographic aspects related to employment, so that the level of investigation of this group, the professional activity carried out, the satisfaction with their main job, the international mobility and the income of this group can be quantified in Spain. 

The study focused on all the doctorate holders resident in Spain, younger than 70, that obtained their degree in a Spanish university, both public or private, between 1990 and 2006. The frame of this statistic operation was a directory of doctorate holders provided to the National Statistic Institute by the University Council, which includes all the persons who have defended a thesis in any Spanish university, according with their electronic databases, which was comprised of approximately 80000 people. Doctors belong to the level 6 of the international classification of education ISCED-97. This level is reserved for tertiary programmes which lead to the award of an advanced research qualification and devoted to advanced study and original research and not based on course-work only.
 
  As for the sampling design, a representative sample was designed for each region at NUTS-2 level\footnote{The NUTS classification (Nomenclature of Territorial Units for Statistics) is a hierarchical system for dividing up the economic territory of the EU for different purposes. (see $http://epp.eurostat.ec.europa.eu/portal/page/nuts\_nomenclature/introduction$)}, using a sampling with equal probabilities. The doctors were grouped according
   to their place of residence, and the selection has been done independently in each
   region by equal probability with random start systematic sampling. A sample of 17000 doctors was selected. The sample has been distributed between the regions assigning the 50\% in an uniform way and the rest proportional to the size of them, measured in number of doctorate holders that have their residence in those regions.
  
  The INE used a questionnaire harmonized at European level, structured in several modules, that can be found at the website of the Institute (http://www.ine.es). As a result of the collection process it was obtained a response rate at the national level of 72\%. 

\begin{figure}[!htb]
 \centering
 \includegraphics[width=1\textwidth]{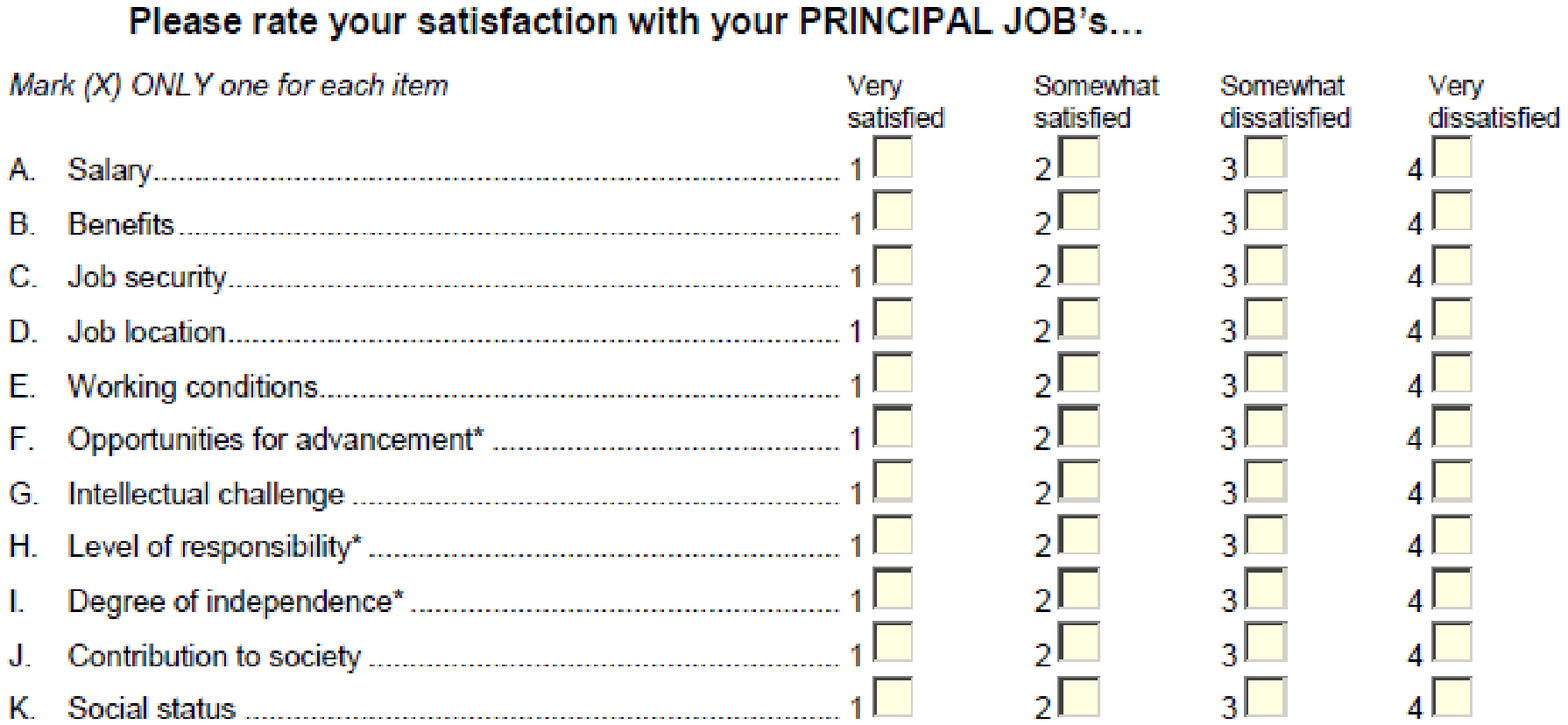}\\[10pt]
 \caption{Pregunta C.6.4 del cuestionario de doctores}
 \label{c64phd}     
\end{figure}

We have 12193 doctorate's answers to the questionnaire in Spain to develop this study. We will focus our attention on the module C (Employment situation) and specifically, in the subsection C.6.4, that tries to find out the level of satisfaction of the doctorate holders in aspects related to their principal job. This question has several points of interest coded on a likert scale from 1 to 4 (see Figure \ref{c64phd}).

Each item will be considered as  ordinal, then we have 11 variables in total.

Using the alternated algorithm to estimate the parameters of the two-dimensional model, we have obtained the indicators in table \ref{fitting} and the factor loadings and communalities in table \ref{factorload}. The percentages of correct classifications are very high for the variables salary and intellectual challenge, presenting a Nagelkerke's $pseudo-R^2$ values
close to one. This makes us thinking that it could be a quasi-separation between categories of the variables, a problem in logistic regression that has been tackled by the estimation method.

Analysing the interpretation of factors using their loadings, the first  has higher
weights for the variables, opportunities for advancement, degree of independence, intellectual challenge, level of responsibility and contribution to society, which are features  associated to research activity.
The second factor presents higher values for Salary, Benefits, Job Security and Working Conditions, all related with economical and working conditions. Then we have two main almost independent factors, the first related to the research activity and the second to the conditions of the job. The Job Location and Social Status variables have similar loadings in both factors. 
\begin{table}[!htb]
\centering\small \caption{Fitting indicators for the eleven variables.}\label{fitting}
\renewcommand{\arraystretch}{0.9}
\begin{tabular}{lcccccccc}\hline
\parbox[c][0.4\height]{10mm}{\centering\smallskip \textbf\tiny{Variable}\smallskip} &%
\parbox[c][0.4\height]{10mm}{\centering\smallskip \textbf\tiny{logLik}\smallskip} &%
\parbox[c][0.5\height]{10mm}{\centering\smallskip \textbf\tiny{Deviance}\smallskip} &%
\parbox[c][1\height]{5mm}{\centering\smallskip \textbf\tiny{df}\smallskip} &%
\parbox[c][1\height]{10mm}{\centering\smallskip \textbf\tiny{p-value}\smallskip} &%
\parbox[c][0.4\height]{5mm}{\centering\smallskip \textbf\tiny{PCC}\smallskip} &%
\parbox[c][0.5\height]{15mm}{\centering\smallskip \textbf\tiny{Nagelkerke}\smallskip}
\\\hline
Salary                    &  -2829.422  &  5658.845 &  2 & 0 & 0.937 & 0.957  \\
Benefits                  &  -8158.998  & 16317.996 &  2 & 0 & 0.799 & 0.826  \\
Job Security              & -12689.867  & 25379.735 &  2 & 0 & 0.582 & 0.151  \\
Job Location              & -11037.486  & 22074.972 &  2 & 0 & 0.603 & 0.098  \\
Working Conditions        & -10072.833  & 20145.666 &  2 & 0 & 0.630 & 0.465  \\
Opp.for Advancement       & -11963.385  & 23926.771 &  2 & 0 & 0.561 & 0.543  \\
Intelectual Challenge     &  -1845.119  &  3690.237 &  2 & 0 & 0.962 & 0.971  \\
Level of responsability   &  -9857.231  & 19714.462 &  2 & 0 & 0.592 & 0.243  \\
Degree of independence    & -10049.245  & 20098.490 &  2 & 0 & 0.609 & 0.386  \\
Contribution to society   &  -9407.924  & 18815.847 &  2 & 0 & 0.623 & 0.247  \\
Social Status             &  -8991.158  & 17982.315 &  2 & 0 & 0.708 & 0.465  \\
\hline
\end{tabular}
\end{table}

\begin{table}[!htb]
\centering\small \caption{Factor loadings and communalities.}\label{factorload}
\renewcommand{\arraystretch}{0.9}
\begin{tabular}{lccc}\hline
\parbox[c][0.4\height]{10mm}{\centering\smallskip \textbf\tiny{Variable}\smallskip} &%
\parbox[c][0.4\height]{10mm}{\centering\smallskip \textbf\tiny{F1}\smallskip} &%
\parbox[c][0.5\height]{10mm}{\centering\smallskip \textbf\tiny{F2}\smallskip} &%
\parbox[c][1\height]{25mm}{\centering\smallskip \textbf\tiny{Communalities}\smallskip} \\\hline
Salary & 0.105  & 0.991   & 0.994 \\
Benefits &  0.109 &  0.986  &  0.984 \\
Job Security &  0.287 &  0.858  &  0.819 \\
Job Location &  0.684 &  0.442  &  0.664 \\
Working Conditions &  0.613 &  0.749  & 0.938 \\
Opportunities for Advancement &  0.876 &  0.403  & 0.930 \\
Intelectual Challenge &  0.988 & -0.137  & 0.995 \\
Level of responsability &  0.902 &  0.173  & 0.843 \\
Degree of independence &  0.911 &  0.275  & 0.906 \\
Contribution to society & 0.922  & 0.018   & 0.851 \\
Social Status & 0.732  & 0.626   & 0.929 \\
\hline
\end{tabular}
\end{table}

\begin{figure}[!htb]
   \centering
   \subfloat[Salary]{\label{fig:iic1} \includegraphics[width=0.33\textwidth]{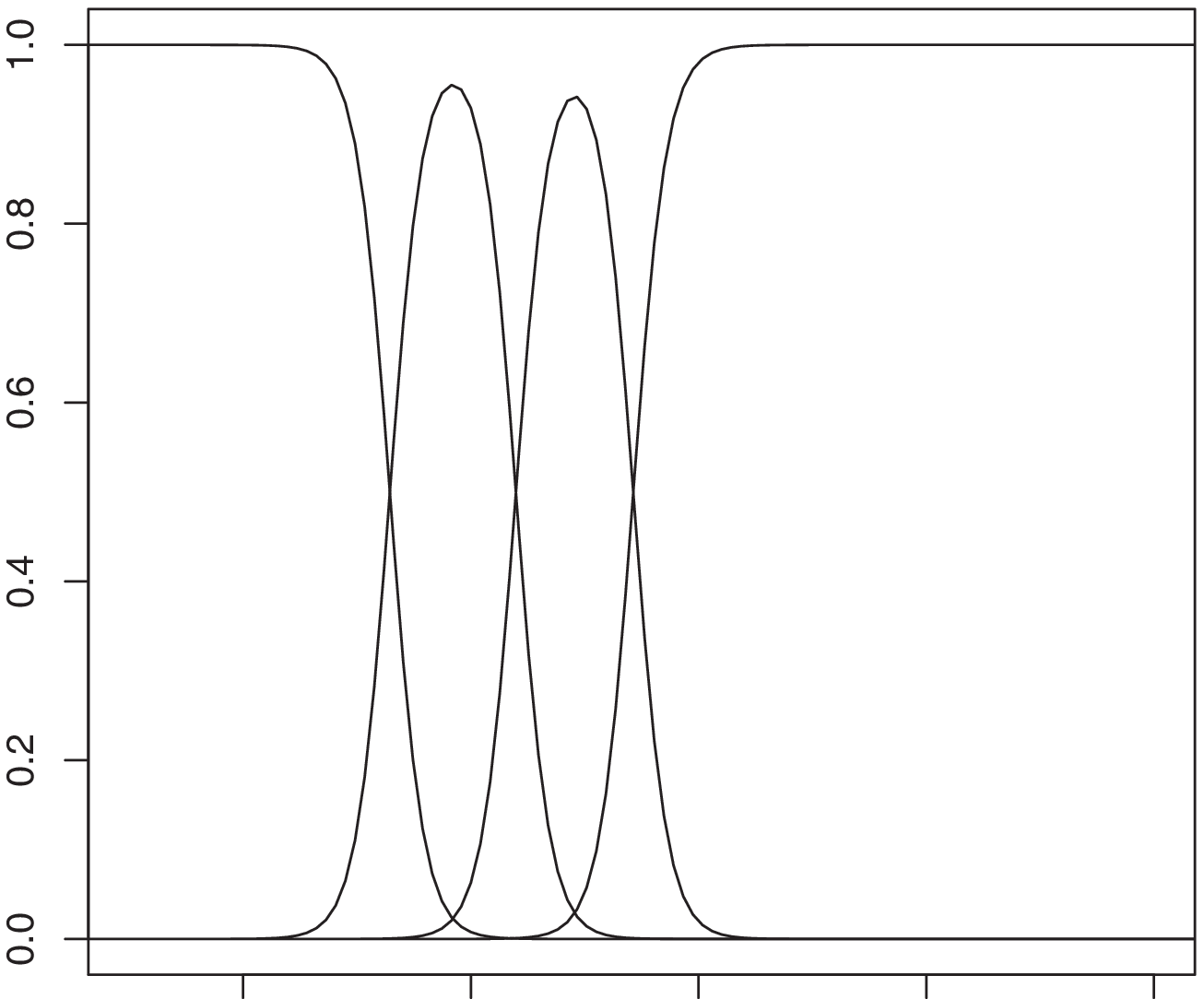}}
   \subfloat[Intellectual challenge]{\label{fig:iic7} \includegraphics[width=0.33\textwidth]{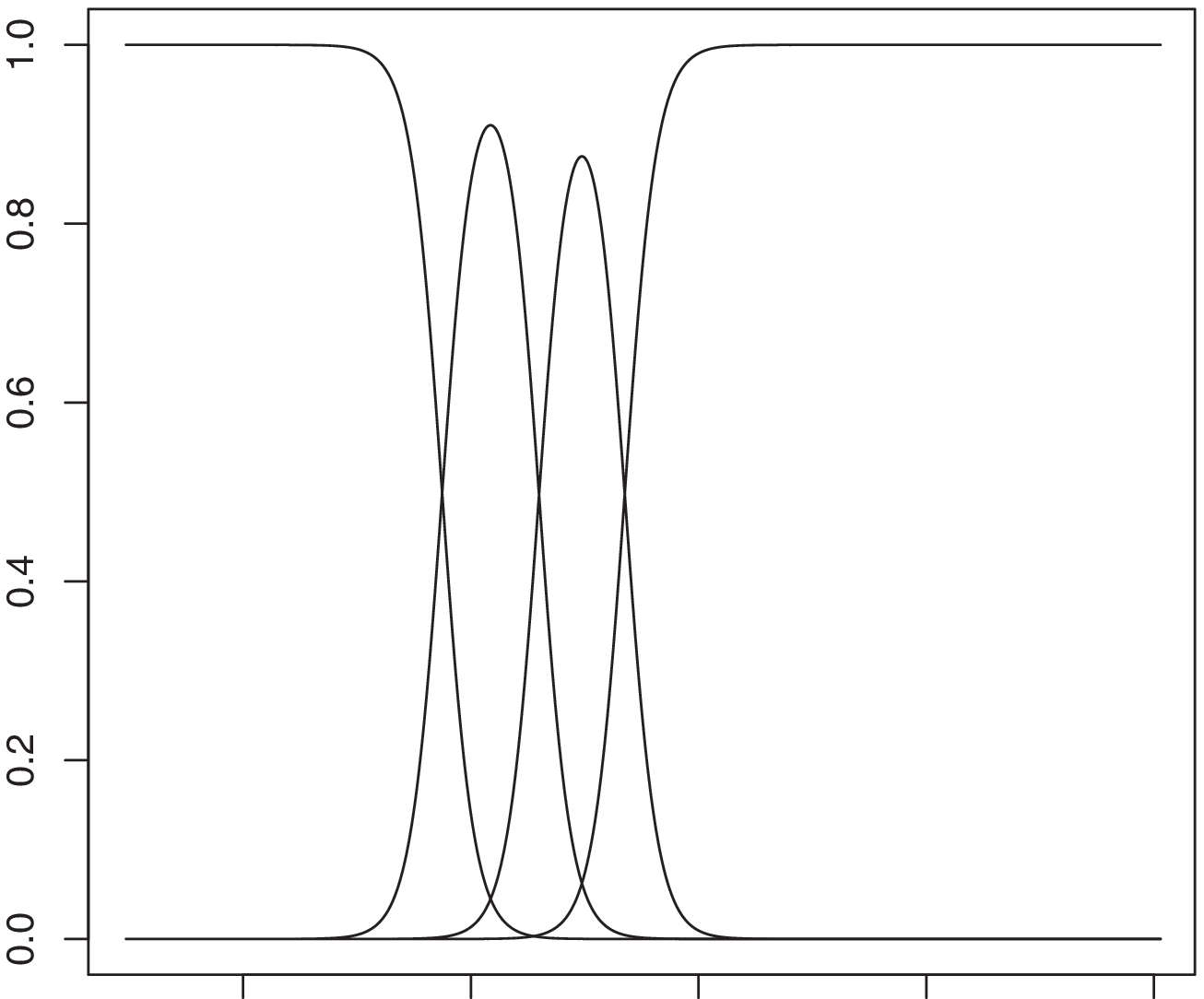}}
   \subfloat[Job security]{\label{fig:iic3} \includegraphics[width=0.33\textwidth]{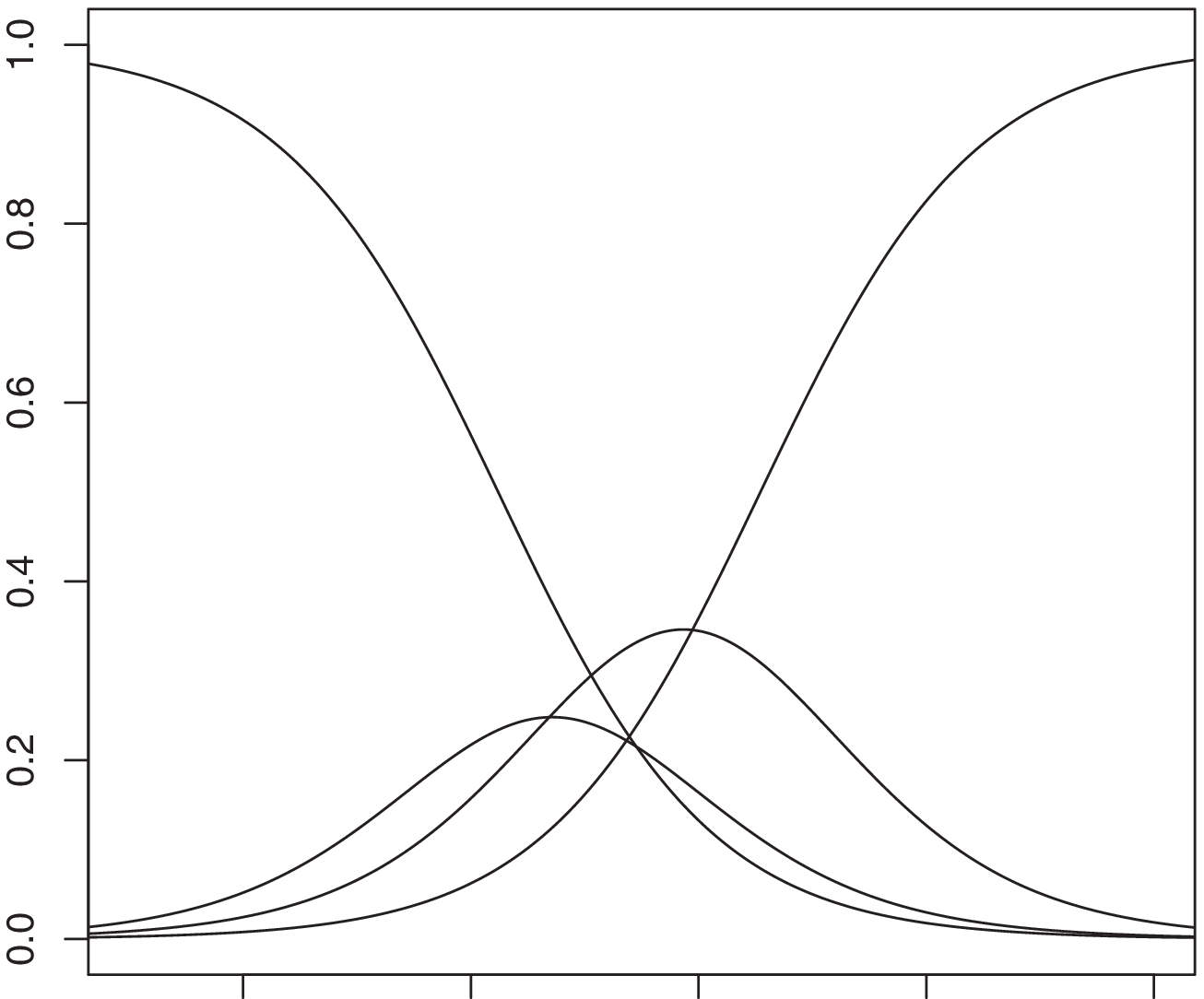}}\\
	[4pt]
   \caption{Curvas de información de los items para cada una de las variables.}
   \label{iics}     
\end{figure}

Item response functions of three of the items can be observed in figure \ref {iics}.

It should be pointed out that in the variable relative to job security, the second category is hidden, which is partially satisfied, concentrating all of the information in the other three categories in such a way, that in this aspect, it appears that either the satisfaction is maximum or is low, which it is in line
with the organization of the Spanish public administration, that concentrates the job of the majority of doctorates.
\begin{figure}[!htb]
 \centering
 \includegraphics[width=0.7\textwidth]{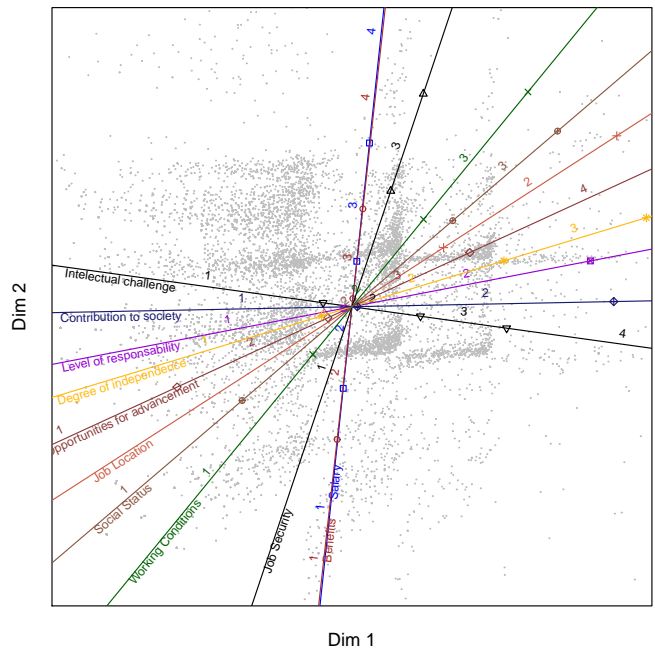}\\[10pt]
 \caption{Ordinal Logistic Biplot. Satisfaction of the doctorate holders with their principal job in Spain.}
 \label{bipord1}     
\end{figure}

The biplot of ordinal data can be seen in figure \ref{bipord1}, in which cut-off points, for each variable, from each of the curves and their projections in the reduced space corresponds to the points scored in each of the lines. In this representation, as mentioned before, the angle between the principal axe with some variables, such as Salary, Benefits, and Job security, is near to $90$\ensuremath{^\circ}, presenting a region in the first quadrant, away from the origin, in which doctorates are very dissatisfied in these aspects and others related with the research activity.

Some of the variables have a similar behavior, such as level of responsibility or contribution to society, in which the points that define the position of each category are distributed in a similar way and their biplot axes present a slope very similar. Although there are groups of variables whose directions are very similar, the position of the categories are quite different from each other within those groups. This happens with opportunities for advancement and degree of independence.

\begin{figure}[!htb]
 \centering
  \subfloat[JobSecurity]{\label{colorindc} \includegraphics[width=0.60\textwidth]{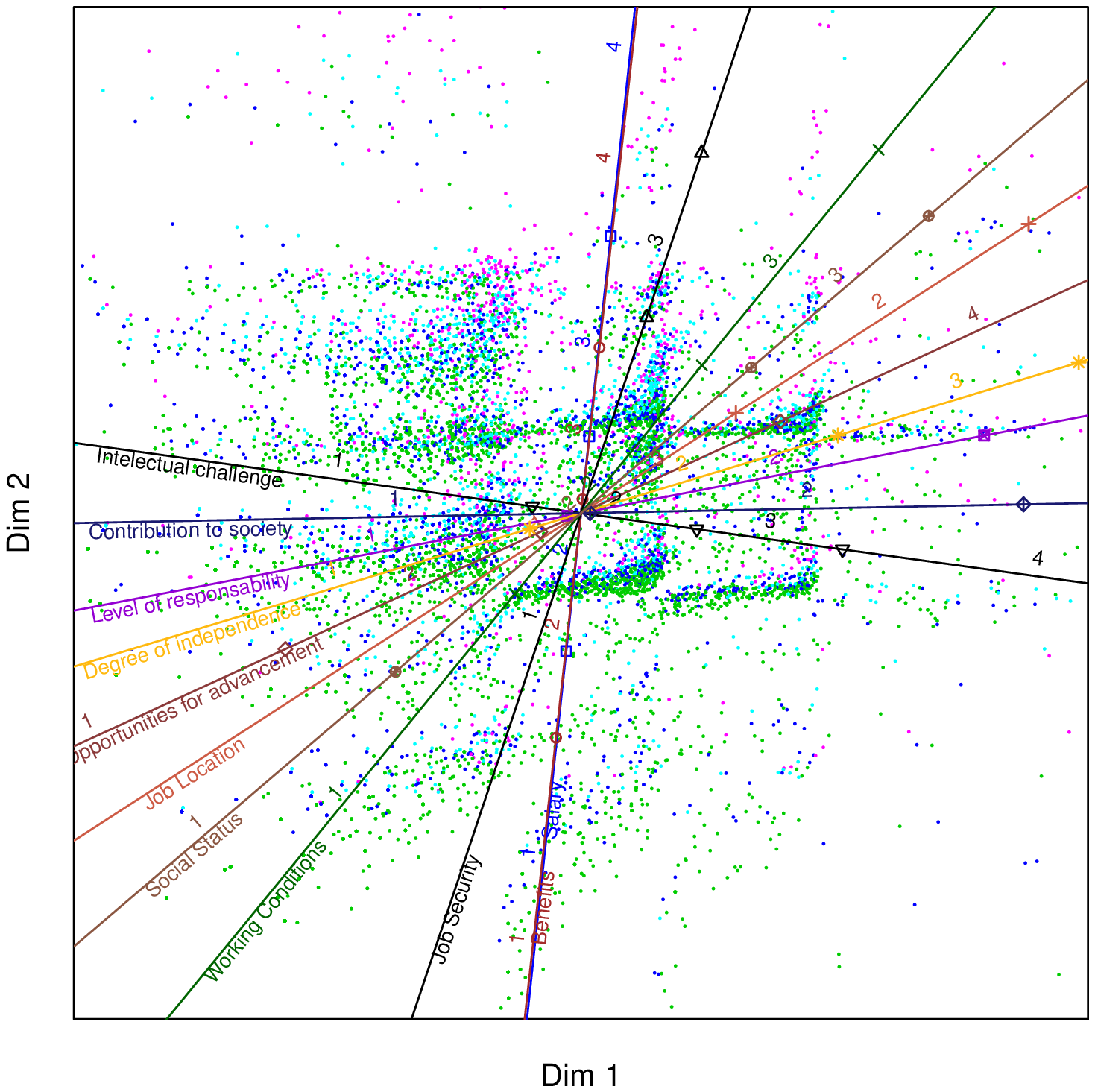}} 
   \subfloat[Intelectual challenge]{\label{colorindb} \includegraphics[width=0.60\textwidth]{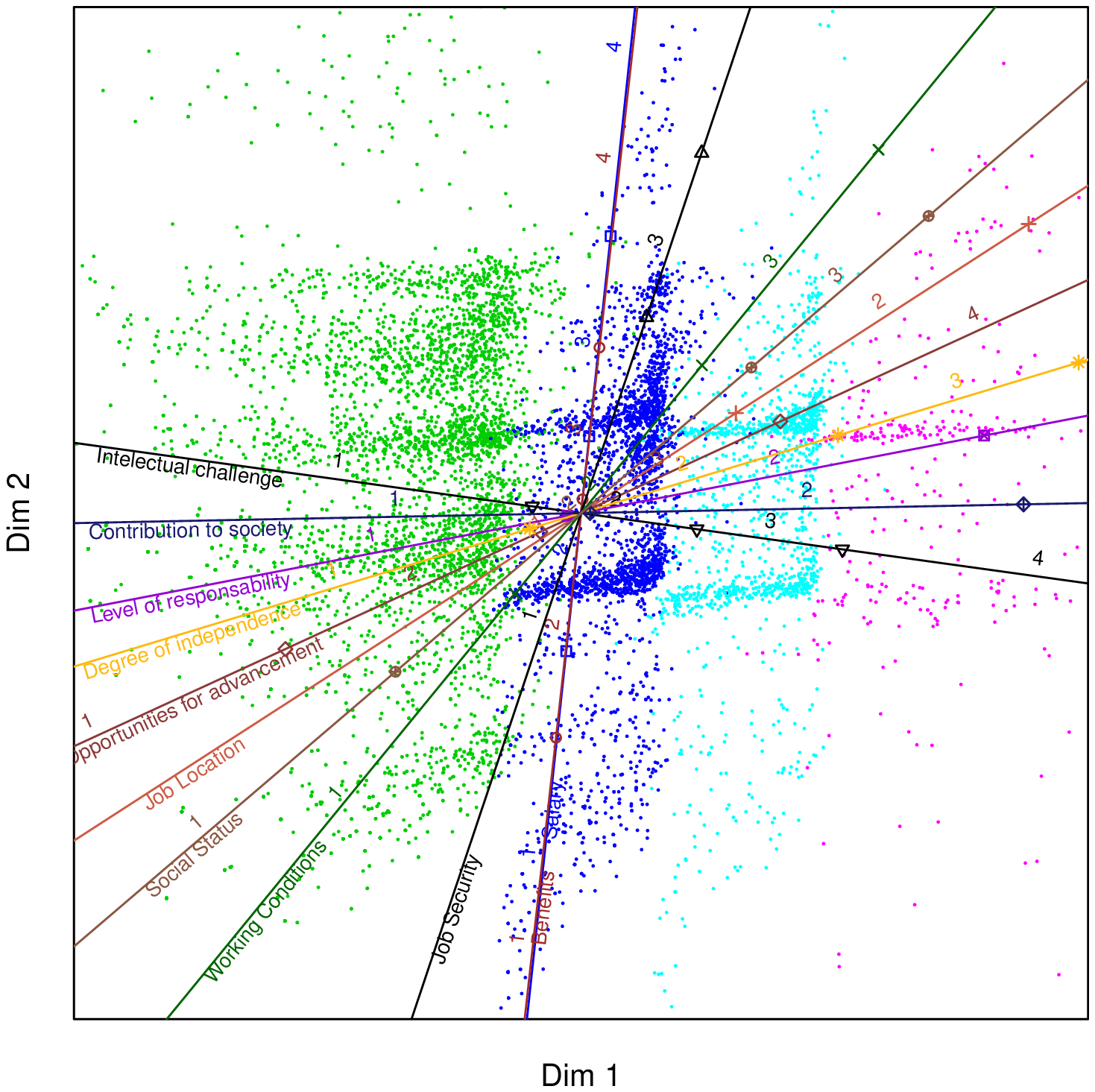}}
   \\[4pt]
   \caption{Coloration according to the category answered by the doctorates in the ordinal logistic biplot.}
   \label{colorationVariables}     
\end{figure}

If we color the individuals according to the answer to the variables represented in the previous curves of information (see figure \ref{colorationVariables}),
the situation of quasi-separation can be appreciated in the Intellectual Challenge variable and not in the  Job Security with an individual behavior more disperse. The Salary also presents this problem of separation, with a graph with horizontal stripes corresponding to each category.
Those variables (Salary and Intellectual Challenge) seem to be important in the interpretation of the information and the understanding of the aspect of the individuals cloud.

Finally, and having in mind the idea of location and characterization more in detail the typology of doctorates in the graph,it has been studied two additional variables, which are the sector where the doctorates work and the region of residence (numbered from 1 to 18, and with a 99 doctorates who are outside Spain). {Figure \ref{centerconvex} shows the convex hulls and centers of sets of points whose category of response is one of the possible for each variable. Regarding to the labour sector, it appears a clear differentiation between higher education sector and the other sectors. For the first one, there are greater intellectual challenge, with a higher degree of independence and opportunities for improvement.
In terms of salary and benefits, the business sector shows a component more attractive for doctorates. As for the residence of doctorates, behaviors are similar, although the perception of the intellectual challenge, the opportunities for improvement, and the degree of independence seems different between the communities of Madrid, Castilla y León, Castilla la Mancha and La Rioja and the rest of Spain.

\begin{figure}[!htb]
 \centering
   \subfloat[Convex Hulls for the sector of employment]{\label{sectorConvexHulls} \includegraphics[width=0.60\textwidth]{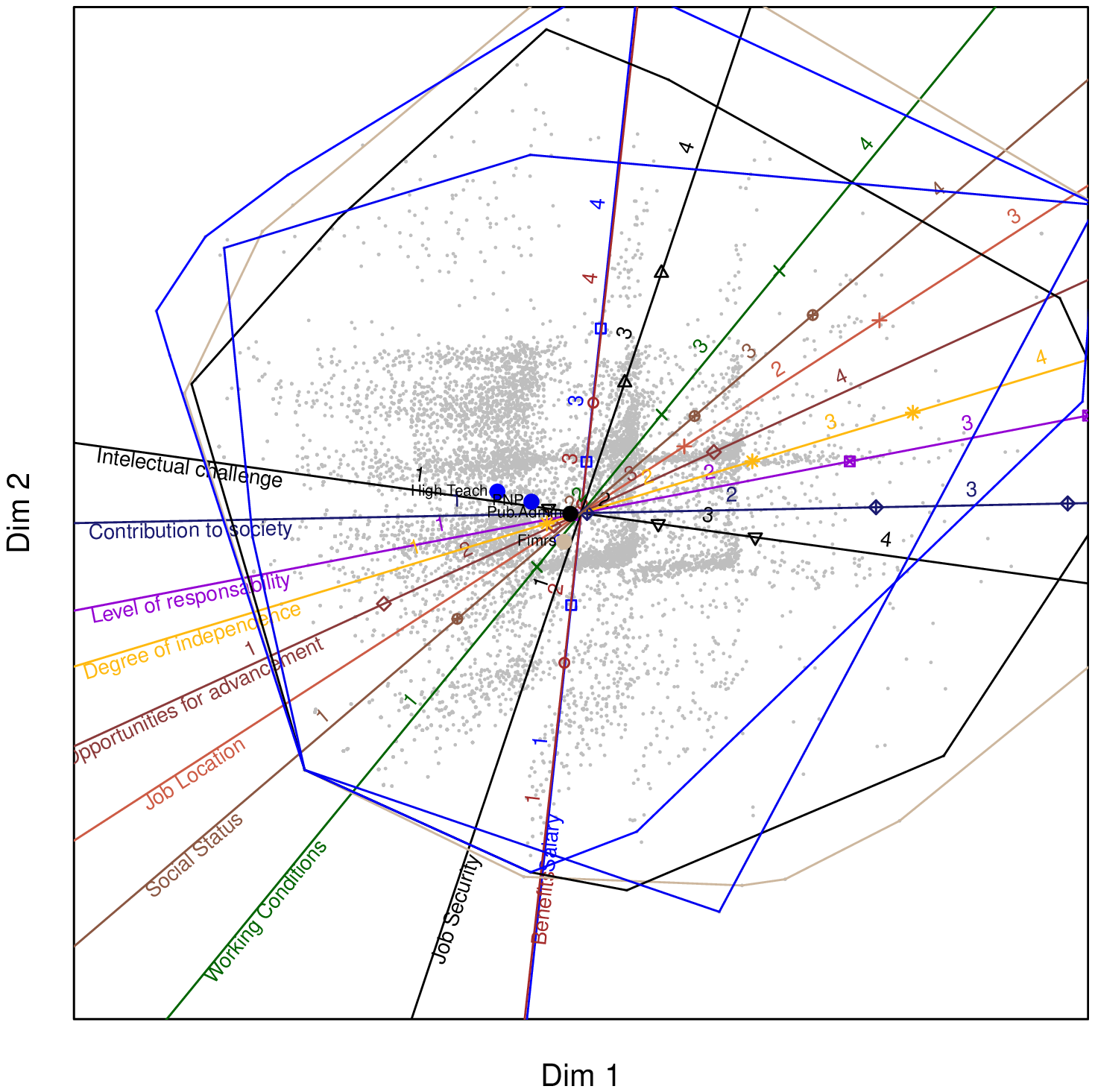}}
   \subfloat[Centers of the sector of employment]{\label{sectorCenters} \includegraphics[width=0.60\textwidth]{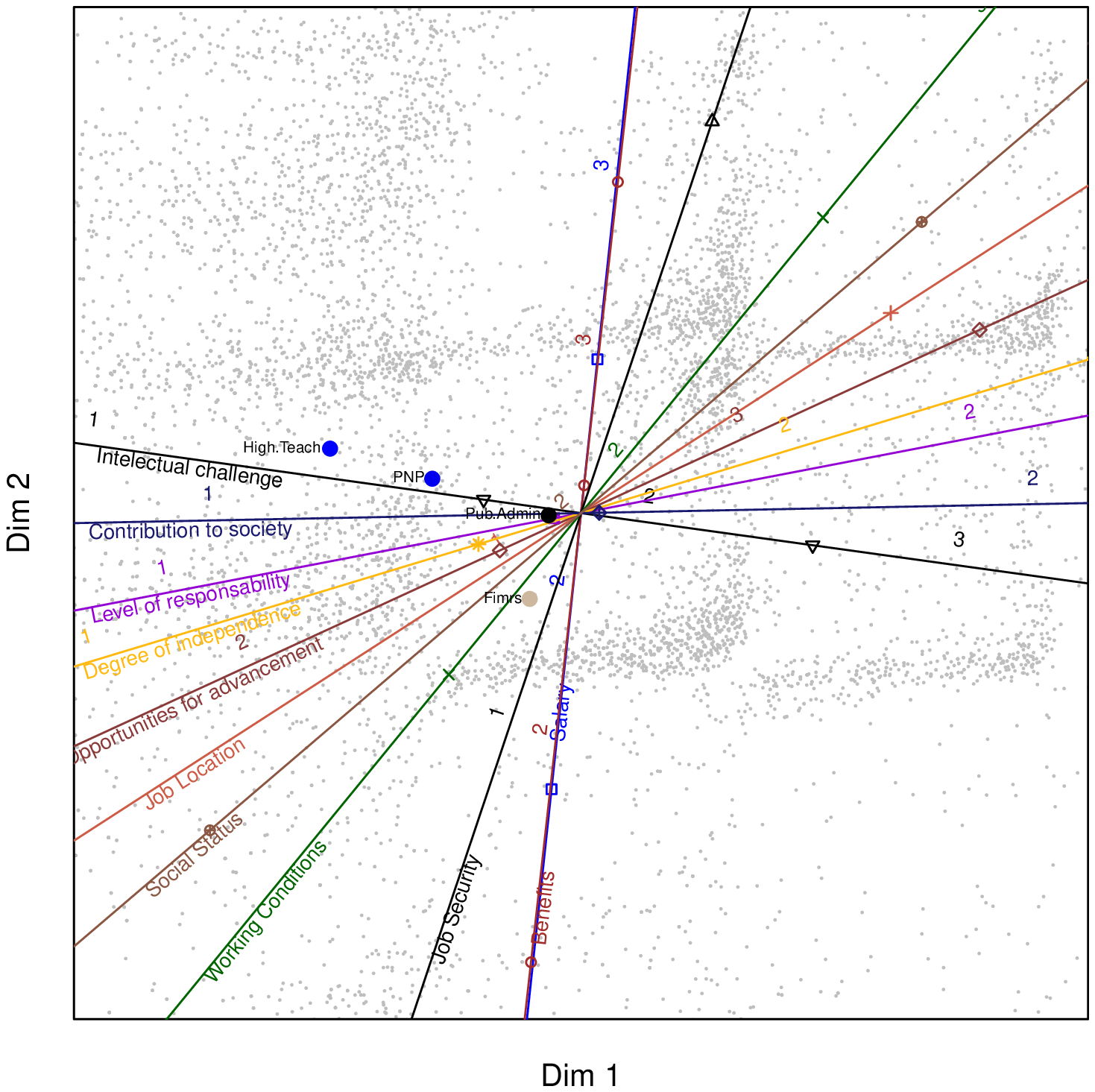}}\\[4pt]
   \subfloat[Convex Hulls for the region of residence]{\label{ccaaConvexHulls} \includegraphics[width=0.60\textwidth]{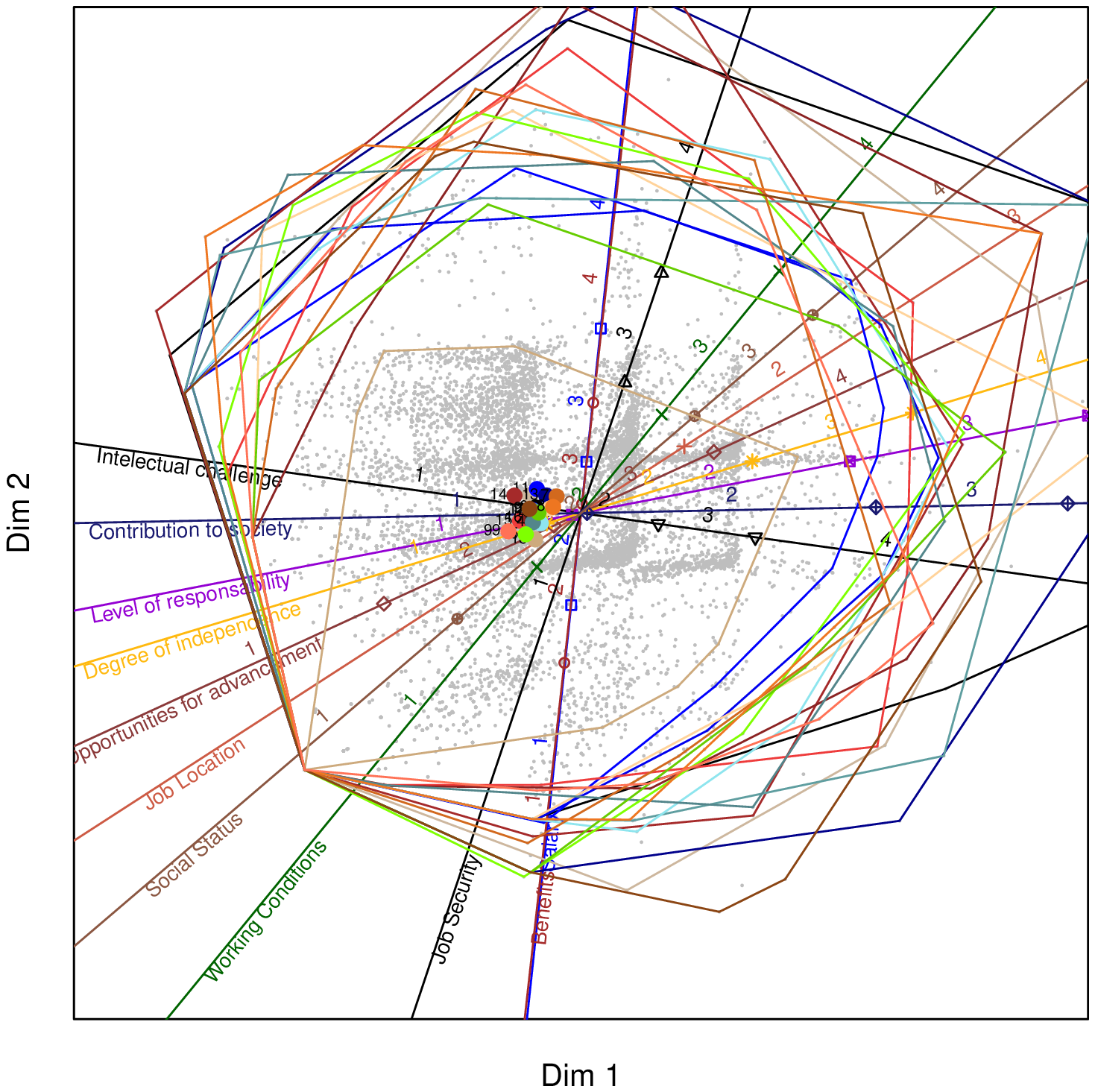}}
   \subfloat[Centers of the region of residence]{\label{ccaaCenters} \includegraphics[width=0.60\textwidth]{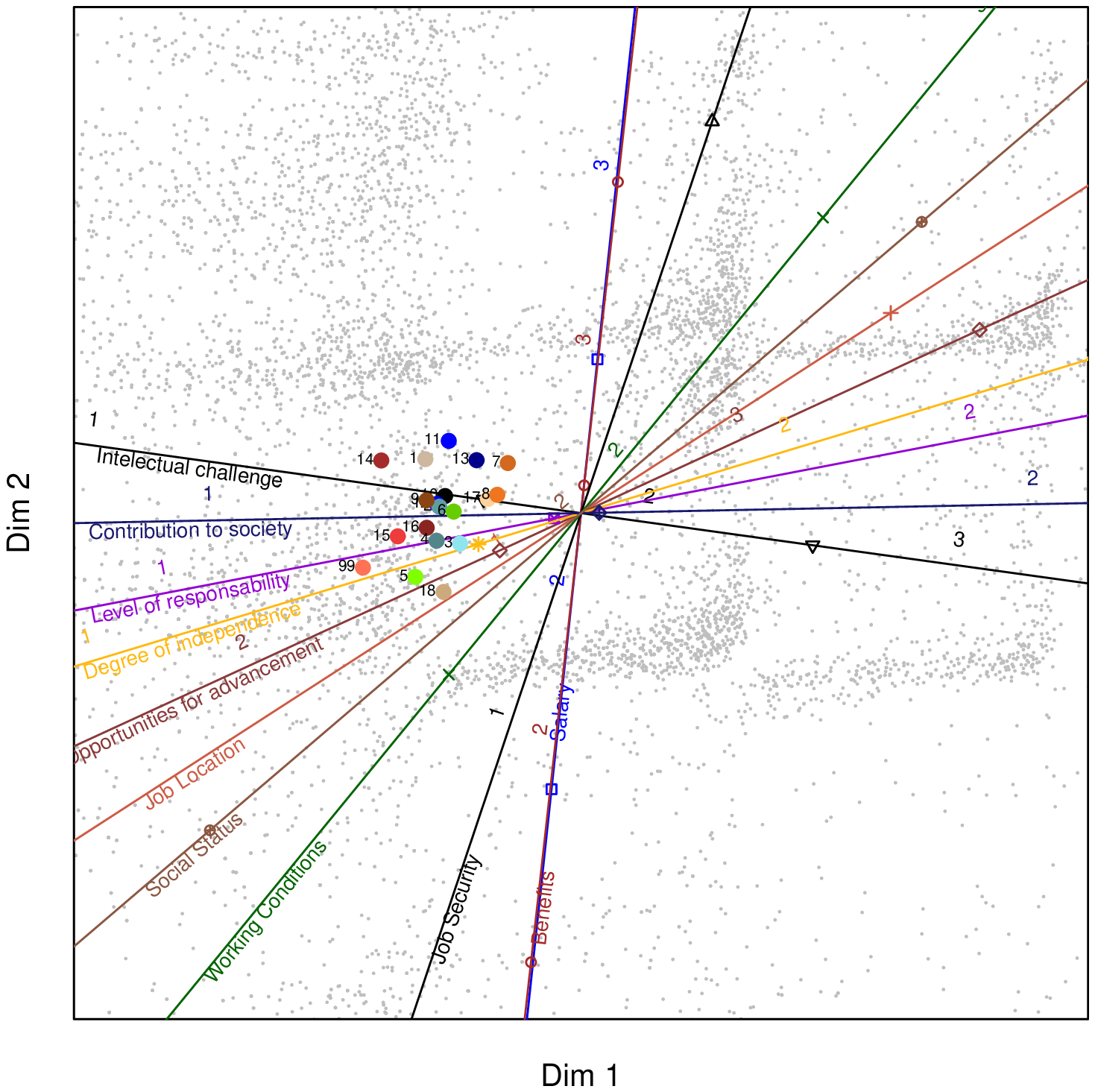}}\\[2pt]      
   \caption{Convex Hulls and centers according to the sector of employment and the region of residence}
   \label{centerconvex}     
\end{figure}

\begin{figure}[!htb]
 \centering
   \subfloat[]{\label{separation} \includegraphics[width=0.5\textwidth]{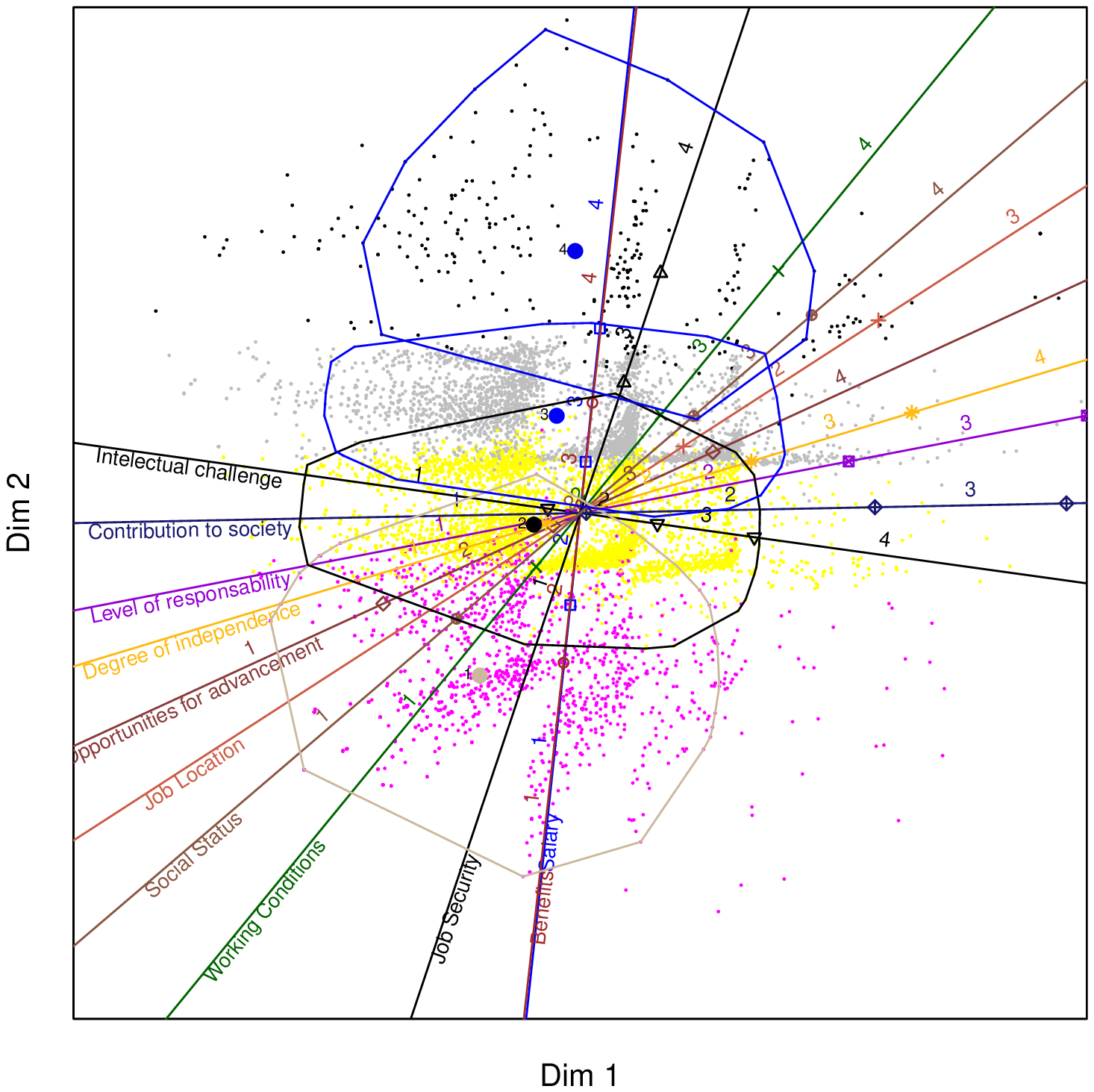}}    \subfloat[]{\label{salary_contour} 
   \includegraphics[width=0.5\textwidth]{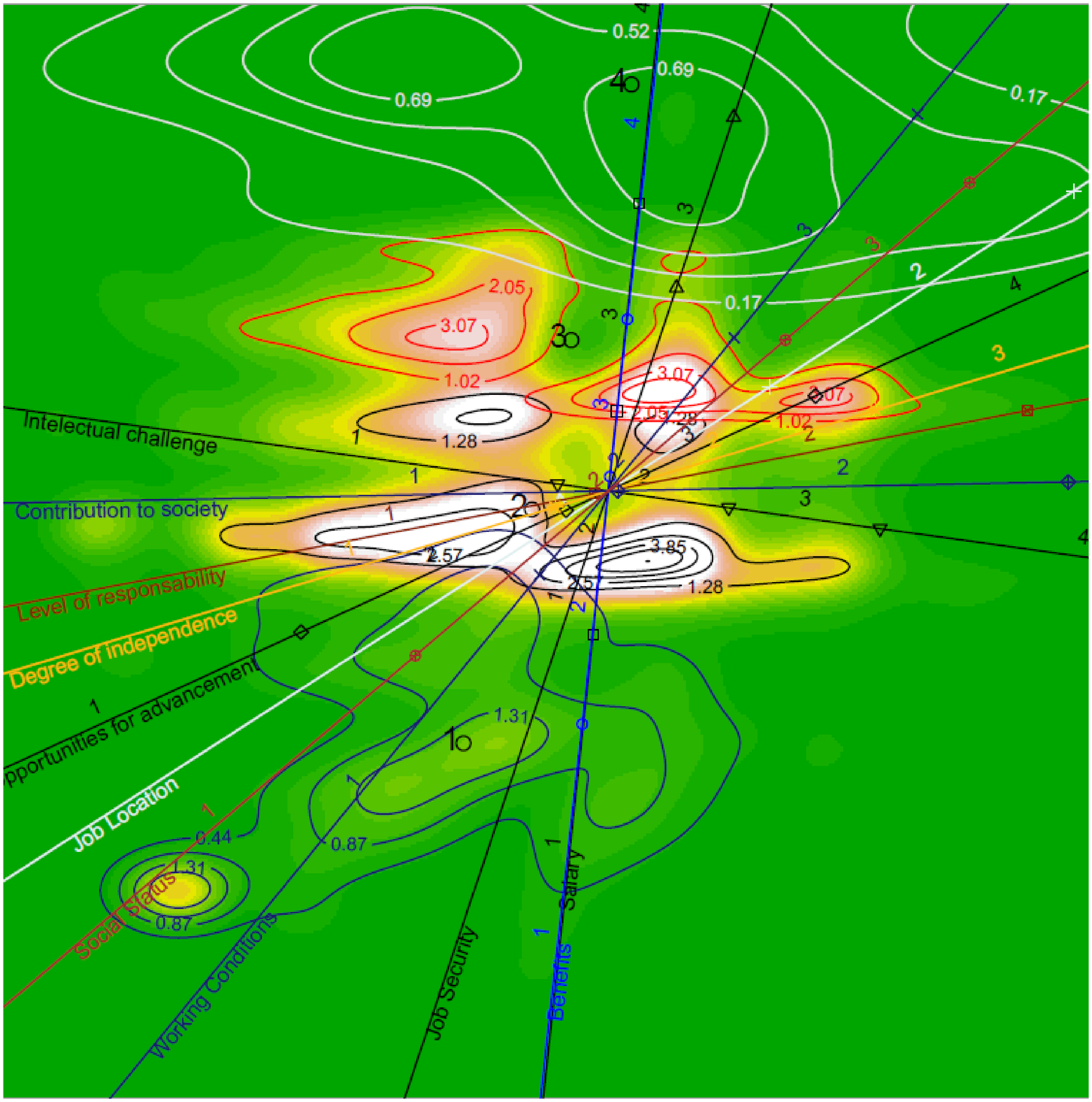}} 
   \caption{Convex hulls and density graph with contour lines for the variable Salary}
   \label{centerconvex}     
\end{figure}

\section{Software Note}\label{soft}
 An R package containing the procedures described by this paper has been developed by the authors \citep{OLB}.

\bibliography{referencesord}
\bibliographystyle{imsart-nameyear}

\end{document}